\documentclass[aps,preprint,prd,nofootinbib,superscriptaddress]{revtex4-1}	
\pdfoutput=1	
\usepackage[utf8]{inputenc}
\usepackage{booktabs}
\usepackage{amssymb,amsmath,amsfonts}
\usepackage{slashed}
\usepackage{indentfirst}
\usepackage[pdftex]{graphicx}
\usepackage{epstopdf}
\usepackage{psfrag}

\usepackage{float,multirow,longtable,rotfloat,supertabular} 
\usepackage{hyperref}
\usepackage{mathrsfs}
\usepackage{url}
\usepackage{appendix}
\usepackage{subfigure}
\usepackage{color}
\usepackage[dvipsnames]{xcolor}
\usepackage{soul}
\usepackage{ulem}
\usepackage{bm}
\usepackage{siunitx}

\allowdisplaybreaks	

\newcommand{\ud}{{\rm{d}}}
\newcommand{\ui}{{\rm{i}}}
\newcommand{\ue}{{\rm{e}}}
\newcommand{\MeV}{\text{ MeV}}
\newcommand{\GeV}{\text{ GeV}}

\def\degree{\ensuremath{{}^{\circ}}}

\graphicspath{{./figures/}}

\begin{document}

\title{Strong decays of excited $ 2^+ $ charmed mesons}

\author{Xiao-Ze Tan}
\affiliation{
	School of Physics, Harbin Institute of Technology, Harbin, 150001, China
}
\author{Tianhong Wang}
\email[corresponding author: ]{thwang@hit.edu.cn}
\affiliation{
	School of Physics, Harbin Institute of Technology, Harbin, 150001, China
}
\author{Yue Jiang}
\affiliation{
	School of Physics, Harbin Institute of Technology, Harbin, 150001, China
}
\author{Qiang Li}
\affiliation{
	School of Physical Science and Technology, Northwestern Polytechnical University, Xi’an 710129, China
}
\author{Lei Huo}
\affiliation{
	School of Physics, Harbin Institute of Technology, Harbin, 150001, China
}
\author{Guo-Li Wang}
\affiliation{
	School of Physics, Harbin Institute of Technology, Harbin, 150001, China
}
\affiliation{
	Department of Physics, Hebei University, Baoding 071002, China
}
\affiliation{
	Hebei Key Laboratory of High-precision Computation and Application of Quantum Field Theory, Baoding 071002, China
}

\begin{abstract}

The new charmed resonance $ D_2^*(3000) $ was observed by the LHCb Collaboration in $ B $ decays. 
In this paper, by assigning it as four possible excited states of the $ 2^+ $ family, we use the instantaneous Bethe-Salpeter method to calculate their Okubo-Zweig-Iizuka-allowed two-body strong decays. The results of $ 1^3F_2 $ and $ 3^3P_2 $ states deviate from the present experimental observation while $ 2^3P_2 $ and $ 2^3F_2 $ are in the error range. Our study also reveals that the widths of these states depend strongly on the masses. Due to the large uncertainties and different mass input, variable models get inconsistent conclusions at the moment.
The analysis in this work can provide essential assistance for future measurements and investigations.

\end{abstract}

\maketitle

\section{Introduction}

During the last decades, many charmed mesons have been found and cataloged \cite{ParticleDataGroup:2020ssz}. With increasing results published, more attention has moved to heavy and excited states. These new states not only enrich the spectrum of $ D $ mesons but also offer us an opportunity to explore their properties from related decays.
In 2013 and 2016, the LHCb Collaboration announced  their observations of several charmed resonances around 3000 MeV, including $ D_J(3000) $, $ D_J^*(3000) $, and $ D_2^*(3000) $ \cite{LHCb:2013jjb,LHCb:2016lxy}. $ D_J(3000) $ and $ D_J^*(3000) $ are observed from $ D^*\pi $ and $ D\pi $ mass
spectrum. Their assignments and strong decays have been widely discussed \cite{Sun:2013qca,Lu:2014zua,Yu:2014dda,Godfrey:2015dva,Gupta:2018zlg,Song:2015fha} . Our previous works prefer they are $ 2P(1^{+ \prime}) $ and $ 2P(0^+) $ states, respectively \cite{Li:2017zng,Tan:2018lao}.  The $D_2^*(3000)$ is observed in the $ B $ decays, whose mass and width are
\begin{equation}
	\begin{aligned}
		M\left(D_{2}^{*}(3000)^{0}\right) &=3214 \pm 29 \pm 33 \pm 36 \MeV, \\
		\Gamma\left(D_{2}^{*}(3000)^{0}\right) &=186 \pm 38 \pm 34 \pm 63 \MeV.
	\end{aligned}
\end{equation}
While the nature of this state is uncertain, it was generally considered as an excited state of the $ 2^+ $ family. 

We summarize some theoretical predictions of the $ 2^+ $ spectrum in Table \ref{spectrum}. $ D_2^*(2460) $, as the ground state $ 1^3P_2 $, was observed and examined by many experiments \cite{ParticleDataGroup:2020ssz}. Our previous work calculated and compared its strong decays with other theoretical studies \cite{Zhang:2016dom, Close:2005se,Zhong:2008kd,Matsuki:2011xp,Godfrey:2015dva}.
For the higher $ 2^3P_2 $, $ 1^3F_2 $, $ 3^3P_2 $, and $ 2^3F_2 $ states their predicted masses range from around 2900 MeV to 3500 MeV. Due to the large uncertainties in the preliminary measurement of $ D_2^*(3000) $, all these four states cannot be ruled out easily from the candidate list. 

\begin{table}[htb]
	\renewcommand\arraystretch{0.9}
	\caption[spectrum]{Theoretical predicted masses of $2^+$ charmed mesons (MeV). }
	\label{spectrum}
	\setlength{\tabcolsep}{5pt} 
	\vspace{0.5em}\centering
	\begin{tabular}{lcccccc}
		\hline
		& Ebert \cite{Ebert:2009ua} & Godfrey \cite{Godfrey:2015dva} & Song \cite{Song:2015fha}\&Wang \cite{Wang:2016krl} & Badalian \cite{Badalian:2020ngz} & Patel \cite{Patel:2021aas} & Ni \cite{Ni:2021pce} \\ \hline
		$1 ^3P_2$ & 2460                     & 2502                          & 2468                                              & 2466                            & 2462                      & 2475                \\
		$2 ^3P_2$ & 3012                     & 2957                          & 2884                                              & 2968                            & 2985                      & 2955                \\
		$1 ^3F_2$ & 3090                     & 3132                          & 3053                                              & 3059                            & 3080                      & 3096                \\
		$3 ^3P_2$ & ...                        & 3353                          & 3234                                              & 3264                            & 3402                      & ...                   \\
		$2 ^3F_2$ & ...                        & 3490                          & 3364                                              & 3430                            & 3494                      & ...                   \\ \hline
	\end{tabular}
\end{table}

To identify the $ D_2^*(3000) $ from these possible assessments, an investigation on their decay behaviors at the experimental mass is necessary.
Usually, the Okubo-Zweig-Iizuka (OZI) allowed strong decays are dominant for charmed mesons. Some efforts about their decays have been made and several theoretical methods were applied, including heavy quark effective theory (HQET) \cite{Wang:2016ewb,Gupta:2018zlg}, the quark pair creation (QPC) model (also named as the $ 3P_0 $ model) \cite{Sun:2013qca,Song:2015fha,Wang:2016krl,Yu:2016mez}, and the chiral quark model \cite{Xiao:2014ura,Ni:2021pce}. Significant discrepancies still exist between different works and the comparison will be discussed in Sec. \ref{sec:results}.

Since the relativistic effects in heavy-light mesons are not negligible, especially for excited states, we use the wave functions obtained by the instantaneous Bethe-Salpeter (BS) approach \cite{Salpeter:1951sz,Salpeter:1952ib} to calculate the OZI allowed two-body strong decays. The BS method has been applied successfully in many previous works \cite{Kim:2003ny,Wang:2007nb,Wang:2009er,Fu:2012zzb,Wang:2012wk,Wang:2013lpa,Li:2016cou,Li:2016efw,Wang:2016mqb,Li:2017sww,Zhou:2019stx} and the relativistic corrections in the heavy flavor mesons are well considered \cite{Geng:2018qrl}. When the light pseudoscalar mesons are involved in the final states, the reduction formula, partially conserved
axial-vector current (PCAC) relation, and low-energy
theorem are used to depict the quark-meson coupling. Since PCAC is inapplicable when the final light meson is a vector, for instance $ \rho $, $ \omega $, and $ K^* $, an effective Lagrangian is adopted instead \cite{Zhong:2009sk,Wang:2016enc}.

The rest of the paper is organized as follows: In Sec. \ref{sec:BS}, we briefly construct BS wave functions of $ 2^+ $ state. Then we derive
the theoretical formalism of the strong decays with the PCAC relation and the effective Lagrangian method. In Sec. \ref{sec:results}, the numerical results and detailed discussions are presented. Finally, we summarize our work in Sec. \ref{sec:summary}.

\section{Theoretical Formalism}
\label{sec:BS}

We first introduce the BS wave functions used in this work. The general form of $ 2^+ $ wave function is constructed as \cite{Wang:2009er,Wang:2013lpa}
\begin{equation}\label{2+BS}
	\begin{aligned}
		\varphi_{2^{+}}\left(q_{\perp}\right) &=\epsilon_{\mu \nu} q_{\perp}^{\mu} q_{\perp}^{\nu}\left[f_{1}\left(q_{\perp}\right)+\frac{\slashed{P}}{M} f_{2}\left(q_{\perp}\right)+\frac{\slashed{q}_{\perp}}{M} f_{3}\left(q_{\perp}\right)+\frac{\slashed{P} q_{\perp}}{M^{2}} f_{4}\left(q_{\perp}\right)\right] \\
		&+M \epsilon_{\mu \nu} \gamma^{\mu} q_{\perp}^{\nu}\left[f_{5}\left(q_{\perp}\right)+\frac{\slashed{P}}{M} f_{6}\left(q_{\perp}\right)+\frac{\slashed{q}_{\perp}}{M} f_{7}\left(q_{\perp}\right)+\frac{\slashed{P} q_{\perp}}{M^{2}} f_{8}\left(q_{\perp}\right)\right],
	\end{aligned}
\end{equation}
where $ M $ and $ P $ are the mass and momentum of the meson, $ q $ is the relative momentum of the quark and antiquark in the meson and $ q_{\perp} $ is denoted as $ q - \frac{P \cdot q}{M^2} P$, $ \epsilon_{\mu \nu} $ is the polarization tensor, and $ f_i $ are the functions of $ q_{\perp} $; their numerical values are obtained by solving the full Salpeter equations \cite{Kim:2003ny,Chang:2004im}. 
Moreover, the wave functions $ f_i $ in Eq.~(\ref{2+BS}) are independent, which are constrained by
\begin{equation}
	\begin{aligned}
		f_{1}\left(q_{\perp}\right) &=\frac{q_{\perp}^{2} f_{3}\left(\omega_{1}+\omega_{2}\right)+2 M^{2} f_{5} \omega_{2}}{M\left(m_{1} \omega_{2}+m_{2} \omega_{1}\right)}, \\
		f_{2}\left(q_{\perp}\right) &=\frac{q_{\perp}^{2} f_{4}\left(\omega_{1}-\omega_{2}\right)+2 M^{2} f_{6} \omega_{2}}{M\left(m_{1} \omega_{2}+m_{2} \omega_{1}\right)}, \\
		f_{7}\left(q_{\perp}\right) &=\frac{M\left(\omega_{1}-\omega_{2}\right)}{m_{1} \omega_{2}+m_{2} \omega_{1}} f_{5}, \\
		f_{8}\left(q_{\perp}\right) &=\frac{M\left(\omega_{1}+\omega_{2}\right)}{m_{1} \omega_{2}+m_{2} \omega_{1}} f_{6},
	\end{aligned}
\end{equation}
where $ m_i $ is the quark/antiquark mass, $ \omega_i = \sqrt{m_i^2-q_{\perp}^2}$, and $ q_{\perp}^2 = -|\vec{q}|^2 $ in the rest frame of the meson.

The review of instantaneous approximation and the action kernel adopted in this paper was given in our previous work \cite{Li:2018eqc}. Due to the approximation, our previous results showed that the predicted mass spectrum for excited heavy-light mesons may not fit the experimental data very well. Thus we follow our previous works\cite{Wang:2012wk,Li:2017sww,Li:2017zng} to use the physical mass as an input parameter to solve the BS wave functions for each state. As we did formerly \cite{Tan:2018lao}, only the dominant positive-energy part of wave functions $ \varphi ^{++} = \Lambda_1^{+} \frac{\slashed{P}}{M} \varphi \frac{\slashed{P}}{M} \Lambda^+_2 $ are kept in the following calculation, which was explained in Ref. \cite{Li:2018eqc} in detail.

\begin{figure}[htbp]
	\centering
	\subfigure[Leading order diagram]{
		\label{fig:feynman1} 
		\includegraphics[width=0.47\textwidth]{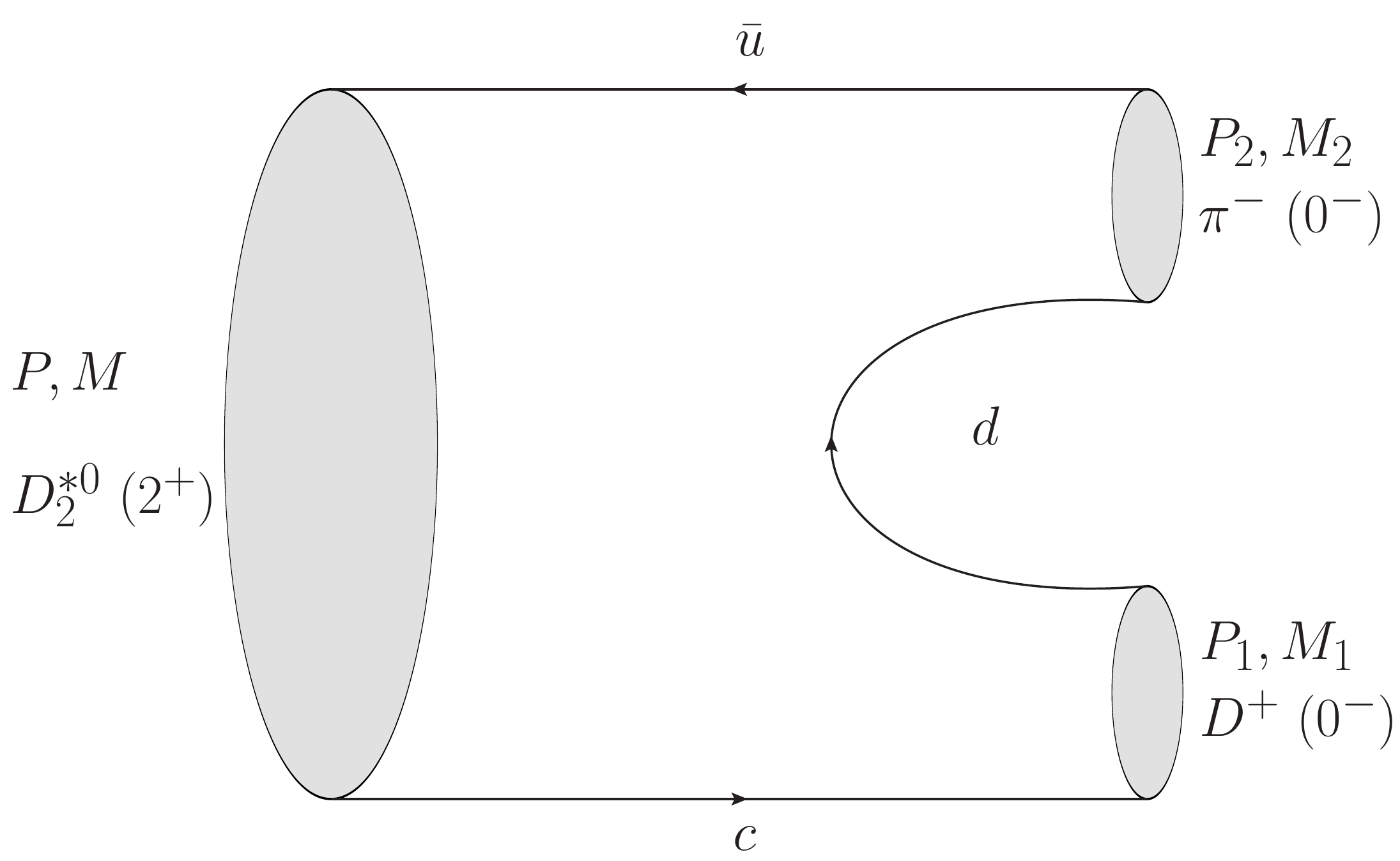}}
	\hspace{0.2cm}
	\subfigure[Approximate diagram]{
		\label{fig:feynman2} 
		\includegraphics[width=0.47\textwidth]{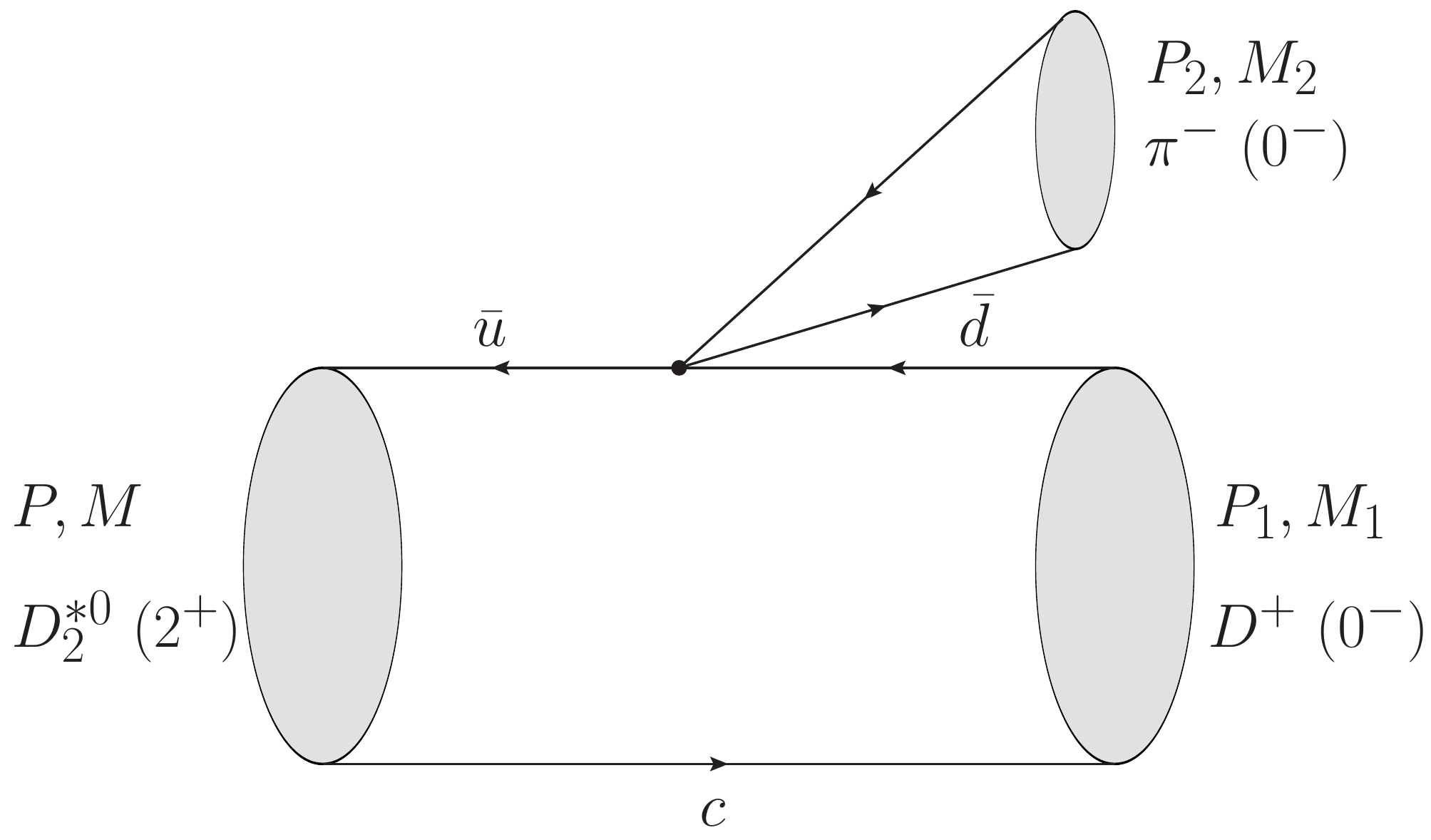}}
	\caption{Feynman diagram of the OZI-allowed two-body strong decay, taking $ D_2^{*0} \to D^+ \pi^- $ for example. }
	\label{fig:feynman}
\end{figure}

Taking $ D_2^* \to D^+ \pi^- $ as an example, the Feynman diagram of two-body strong decay is shown in Fig. \ref{fig:feynman}. 
By using the reduction formula, the corresponding transition matrix element has the form 
\begin{equation}\label{matrix}
	\begin{aligned}
		&\langle D^+(P_{1})\pi ^-(P_{2})\left| D_2^{*0}(P)\right. \rangle =\int \ud ^4x \ue ^{ \ui P_{2}\cdot x} (M_{\pi}^2 -P_{2}^2)\langle D^+(P_{1})\left| \phi _{\pi}(x)\right| D_2^{*0}(P) \rangle ,
	\end{aligned}
\end{equation}
in which $\phi_{\pi}$ is the light pseudoscalar meson field. By using the PCAC relation, the field is expressed as  
\begin{equation}\label{axialvector}
	\phi_{\pi}(x)=\frac{1}{M_{\pi}^2 f_{\pi}}\partial ^{\mu}(\overline{u}\gamma _{\mu} \gamma_{5}d),
\end{equation}
where $f_{\pi}$ is the decay constant of $ \pi $. Then, by using the low-energy approximation, the transition amplitude in momentum space can be derived as \cite{Chang:2005sd} 
\begin{equation}
	\begin{aligned}
		\langle D^+(P_{1}) & \pi ^-(P_{2})\left| D_2^{*0}(P)\right. \rangle \\
		&\approx -\ui \frac{P_{2}^{\mu}}{f_{\pi}} (2\pi)^4 \delta ^4 (P -P_{1}-P_{2}) \langle D^+(P_{1})\left| \overline{u}\gamma _{\mu} \gamma_{5}d \right| D_2^{*0}(P)\rangle  .
	\end{aligned}
\end{equation}

Besides the PCAC relation, the same transition amplitude can be obtained by introducing an effective Lagrangian \cite{Zhong:2009sk,Wang:2016enc,Li:2017zng}
\begin{equation}
	\mathcal{L}_{qqP}=\frac{g}{\sqrt{2}f_h} \bar{q}_{i} \gamma^{\xi} \gamma^5 q_j \partial_{\xi}\phi_{ij},
\end{equation}
where
\begin{equation}
	\phi_{ij}=\sqrt{2}
	\left[
	\begin{matrix}
		\frac{1}{\sqrt{2}} \pi^0 + \frac{1}{\sqrt{6}}\eta & \pi^+ & K^+ \\
		\pi^-	&	\-\frac{1}{\sqrt{2}}\pi^0 + \frac{1}{\sqrt{6}}\eta	&	K^0	\\
		K^-	&	K^0	&	-\frac{2}{\sqrt{6}}\eta
	\end{matrix}
	\right]
\end{equation}
denotes the chiral field of the light pseudoscalar meson, $g$ is the quark-meson coupling constant and $f_h$ is the decay constant.

For further numerical calculation, in the Mandelstam formalism \cite{Mandelstam:1955sd}, we can write the hadronic transition amplitude as the overlapping integration over the relativistic wave functions of the initial and final mesons \cite{Chang:2005sd} 
\begin{equation}\label{feynman:amp}
	\begin{split}
		\mathcal{M}=& -\ui \frac{P_{2}^{\mu}}{f_{\pi}}\langle D^+(P_{1})\left| \overline{u}\gamma _{\mu} \gamma_{5}d \right| D_2^{*0}(P)\rangle \\
		=&-\ui \frac{P_{2}^{\mu}}{f_{\pi}} \int \frac{\ud ^3 q}{(2 \pi)^3} \mathrm{Tr}\left[\overline{\varphi} _{P_{1}}^{++}({q}_{1\perp})\frac{\slashed{P}}{M}\varphi_{P}^{++}({q}_{\perp})\gamma _{\mu}\gamma _{5}\right],
	\end{split}
\end{equation}
where $ \varphi^{++}$ and $ \overline{\varphi}^{++}$ are the positive part of BS wave function and its Dirac adjoint form, 
and the quark-antiquark relative momenta in the initial and final meson have the relation  $q_{1}=q-\frac{m_c}{m_c+m_d}P_{1}$.

When the final light meson is $\eta$ or $\eta '$ instead of $ \pi $, we also consider the $\eta-\eta '$ mixing \cite{ParticleDataGroup:2020ssz}
\begin{equation}
	\left(
	\begin{array}{c}
		\eta \\
		\eta '
	\end{array}
	\right)
	=
	R^T(\theta_{\eta})
	\left(
	\begin{array}{c}
		\eta_8 \\
		\eta_1
	\end{array}
	\right),
\end{equation}
where 
\begin{equation}
	\begin{aligned}
		&\eta_8= \frac{1}{\sqrt{6}} (u\bar{u}+d\bar{d}-2s\bar{s}) , \\
		&\eta_1= \frac{1}{\sqrt{3}} (u\bar{u}+d\bar{d}+s\bar{s}) .
	\end{aligned}
\end{equation}
$ R(\theta) $ or $ R^T(\theta) $ is the mixing matrix defined as 
\begin{equation}
	R(\theta) = 
	\left(
	\begin{array}{cc}
		\cos \theta  & \sin \theta \\
		-\sin \theta & \cos \theta \\
	\end{array}
	\right), \quad
	R^T(\theta) = 
	\left(
	\begin{array}{cc}
		\cos \theta & -\sin \theta \\
		\sin \theta & \cos \theta
	\end{array}
	\right) .
\end{equation}

We choose the mixing angle $\theta_{\eta} = -11.3^ \circ$ \cite{ParticleDataGroup:2020ssz} in this work. By extracting the coefficient of mixing, the transition amplitudes with $ \eta/\eta ' $ involved are
\begin{equation}
	\begin{split}
		\mathcal{M}(\eta)=- \ui P_{2}^{\mu} M^2_{\eta} \left(
		\frac{\cos \theta_\eta}{\sqrt{6}f_{\eta_8}M^2_{\eta_8}}-\frac{\sin \theta_\eta}{\sqrt{3}f_{\eta_1}M^2_{\eta_1}}
		\right) \langle D^{(*)}(P_{1})\left| \overline{u}\gamma _{\mu} \gamma_{5}u \right| D_2^*(P)\rangle, \\
		\mathcal{M}(\eta ')=- \ui P_{2}^{\mu} M^2_{\eta '}\left(
		\frac{\sin \theta_\eta}{\sqrt{6}f_{\eta_8}M^2_{\eta_8}}+\frac{\cos \theta_\eta}{\sqrt{3}f_{\eta_1}M^2_{\eta_1}}
		\right) \langle D^{(*)}(P_{1})\left| \overline{u}\gamma _{\mu} \gamma_{5}u \right| D_2^*(P)\rangle.
	\end{split}
\end{equation}

When the final light meson is a vector, such as $ \rho $ or $ \omega $, the PCAC rule is not valid. Thus we adopt the effective Lagrangian method to derive the transition amplitude. The effective Lagrangian of light vector meson is given by \cite{Zhong:2009sk,Wang:2016enc,Li:2017zng} 
\begin{equation}\label{efl}
	\mathcal{L}_{qqV}=\sum_j \bar{q}_j(a \gamma_{\mu}+\frac{\ui b}{2 m_{j}}\sigma_{\mu \nu}P_{2}^\nu )V^{\mu}q_j,
\end{equation}
where $ m_j $ is the constitute quark mass (we take approximation $ m_j=(m_q+m_{\bar{q}})/2 $ in this work), $ \sigma_{\mu \nu} = \frac{\ui}{2}\left[\gamma_{\mu}, \gamma_{\nu}\right] $; $V^{\mu}$ is the light vector meson field, and the parameters $a=-3$ and $b=2$ denote the vector and tensor coupling strength \cite{Wang:2016enc}, respectively. Then we use Eq.~(\ref{efl}) directly to get the vertex of the light vector and reach the transition amplitude 
\begin{equation}\label{feynman:ampefl}
	\begin{split}
		\mathcal{M}=-\ui \int \frac{\ud ^3 q}{(2 \pi)^3} \mathrm{Tr}\left[\overline{\varphi} _{P_{1}}^{++}({q}_{1\perp})\frac{\slashed{P}}{M}\varphi_{P}^{++}({q}_{\perp})(a \gamma_{\mu}+\frac{\ui b}{2 m_j}\sigma_{\mu \nu}P_{2}^\nu )\varepsilon_{2}^{\mu}\right] .
	\end{split}
\end{equation}

In the possible strong decays of $ D_2^*(3000) $, $ 1^+ $ and $ 2^- $ states of  $ D $ mesons are also involved in the final state. In the heavy quark limit($m_Q\to \infty$), the coupling of spin $ S $ and orbital angular
momentum $ L $ no longer describes the physical states well for these heavy-light states. The $ ^1P_1 $ - $ ^3P_1 $ and $ ^1D_2 $ - $ ^3D_2 $ mixing are needed. We take the total angular momentum of the light quark in the mesons $\vec{j}_l=\vec{L}+\vec{s}_q$ ($ s_q $ is the light-quark spin) to identify the physical doublet. The mixing relations are given by \cite{Barnes:2005pb,Ebert:2009ua,Matsuki:2010zy,Li:2016efw}
\begin{equation}
	\begin{aligned}
		\left(
	\begin{array}{c}
		|J^P=1^+,j_l={3}/{2} \rangle \\
		|J^P=1^+,j_l={1}/{2} \rangle \\
	\end{array}
	\right)
	=
	R(\theta_{1P})
	\left(
	\begin{array}{c}
		|{^1}P_1 \rangle \\
		|{^3}P_1 \rangle \\
	\end{array}
	\right), \\
		\left(
	\begin{array}{c}
		|J^P=2^-,j_l={5}/{2} \rangle \\
		|J^P=2^-,j_l={3}/{2} \rangle \\
	\end{array}
	\right)
	=
	R(\theta_{1D})
	\left(
	\begin{array}{c}
		|{^1}D_2 \rangle \\
		|{^3}D_2 \rangle \\
	\end{array}
	\right),
	\end{aligned}
\end{equation}
where the ideal mixing angles $ \theta_{1P}=\arctan(\sqrt{1/2}) \approx 35.3 \degree$ and $ \theta_{1D}=\arctan(\sqrt{2/3}) \approx 39.2\degree $ in the heavy-quark limit are adopted. We notice that varying mixing angles could make difference to the decay widths. The dependence of corresponding partial widths on mixing angle will be discussed later.

In our calculation, the Salpeter equations for $^1P_1$ and $^3P_1$ ($ ^1D_2 $ and $ ^3D_2 $) states are solved individually and the state messes are mixing by the physical masses values
\begin{equation}
	\begin{aligned}
		\left(
		\begin{array}{cc}
			m_{{^1} P_{1}}^{2} & \delta \\
			\delta	& m_{{^3} P_{1}}^{2}
		\end{array}
		\right) = 
		R^T(\theta_{1P})
		\left(
		\begin{array}{cc}
			m_{3 / 2}^{2} & 0 \\
			0	& m_{1 / 2}^{2}
		\end{array}
		\right)
		R(\theta_{1P}), \\
		\left(
		\begin{array}{cc}
			m_{{^1} D_{2}}^{2} & \delta \\
			\delta	& m_{{^3} D_{2}}^{2}
		\end{array}
		\right) = 
		R^T(\theta_{1D})
		\left(
		\begin{array}{cc}
			m_{5 / 2}^{2} & 0 \\
			0	& m_{3 / 2}^{2}
		\end{array}
		\right)
		R(\theta_{1D}) .
	\end{aligned}
\end{equation}

According to Refs. \cite{Cheng:2006dm,Ju:2014oha,Li:2018eqc,Li:2016efw}, we specify $ D_1(2420) / D_{s1}(2536) $ as $| j_l={3}/{2} \rangle $ and $ D_1(2430) / D_{s1}(2460) $ as $| j_l={1}/{2} \rangle $ for $ 1^+ $ state, $ D_2(2740)$ as $| j_l={5}/{2} \rangle $ and $ D_2(2780) $ as $| j_l={3}/{2} \rangle $ for $ 2^- $ state within this paper. 

By performing the integration and trace in Eqs.~(\ref{feynman:amp}) and (\ref{feynman:ampefl}), the amplitudes of all possible channels within the present study can be simplified as
\begin{align}\label{amp}
	&\mathcal{M}_{(2^+\to 0^- 0^-)}=  \frac{1}{f_{P_2}} t_1 \epsilon_{\mu \nu}  P_1^{\mu} P_1^{\nu} , 
	\nonumber \\
	&\mathcal{M}_{(2^+\to 0^- 1^-)}= t_2 \epsilon_{\mu \nu} \epsilon_{2  \alpha} P_{\beta} P_{1\gamma} P_{1}^{\mu}   \varepsilon^{\nu \alpha \beta \gamma} ,
	\nonumber \\
	&\mathcal{M}_{(2^+\to 0^+ 1^-)}= \epsilon_{\mu \nu } \epsilon_{2\alpha} \left( t_3 P_1^{\mu} P_1^{\nu} P^{\alpha} + t_4 P_1^{\nu}  g^{\mu \alpha} \right) ,
	\nonumber \\
	&\mathcal{M}_{(2^+\to 1^-  0^-)}=  \frac{\ui}{f_{P_2}} t_5 \epsilon_{\mu \nu} \epsilon_{1 \alpha} P_{\beta} P_{1 \gamma} P_1^{\mu}   \varepsilon^{\nu \alpha \beta \gamma} ,
	\nonumber \\
	&\mathcal{M}_{(2^+\to 1^- 1^-)}=  \epsilon_{\mu \nu} \epsilon_{1\alpha} \epsilon_{2\beta} \left[ P_1^{\mu} \left( t_6 P_1^{\nu} P^{\alpha} P^{\beta} + t_7 P_1^{\nu} g^{\alpha \beta} + t_8 P^{\alpha} g^{\nu \beta}  + t_9 P^{\beta} g^{\nu \alpha} \right) + t_{10} g^{\mu \alpha} g^{\nu \beta} \right] ,
	\nonumber \\
	&\mathcal{M}_{(2^+\to 1^{+(\prime)}  0^-)}=  \frac{\ui}{f_{P_2}} \epsilon_{ \mu \nu} \epsilon_{1 \alpha} \left( t_{11}^{(\prime)} P_1^{\mu} P_1^{\nu} P^{\alpha} + t_{12}^{(\prime)} P_1^{\mu} g^{\nu \alpha} \right) ,
	\nonumber \\
	&\mathcal{M}_{(2^+\to 1^{+(\prime)}  1^-)}= \epsilon_{\mu \nu} \epsilon_{1\alpha} \epsilon_{2\beta} \left[ P_1^{\mu} \varepsilon^{ \nu \alpha \beta \gamma} \left( t_{13}^{(\prime)} P_{\gamma} + t_{14}^{(\prime)} P_{1\gamma} \right) + t_{15}^{(\prime)} P_1^{\mu} P^{\beta} P_{\gamma} P_{1\delta} \varepsilon^{\nu \alpha \gamma \delta } + t_{16}^{(\prime)} P_{\gamma} P_{1\delta} g^{\mu \alpha} \varepsilon^{\nu \beta \gamma \delta} \right. 
	\nonumber \\ 
	& \qquad \qquad  \qquad \quad + \left. t_{17}^{(\prime)} P_1^{\mu} P_1^{\nu} P_{\gamma} P_{1\delta} \varepsilon^{\alpha \beta \gamma \delta} \right] ,
	\nonumber \\
	&\mathcal{M}_{(2^+\to 2^+ 0^-)}=  \frac{\ui}{f_{P_2}} \epsilon_{\mu \nu} \epsilon_{1\alpha \beta} P_{\gamma} P_{1\delta} \varepsilon^{\nu \beta \gamma \delta}  \left( t_{18} P_1^{\mu} P^{\alpha} + 
	 t_{19}  g^{\mu \alpha}    \right) ,
	\nonumber \\
	&\mathcal{M}_{(2^+\to 2^{-(\prime)} 0^-)}= \frac{1}{f_{P_2}} \epsilon_{\mu \nu} \epsilon_{1 \alpha \beta} \left[ P_1^{\mu} P^{\alpha} \left( t_{20}^{(\prime)} P_1^{\nu} P^{\beta} + t_{21}^{(\prime)} g^{\nu \beta} \right) + t_{22}^{(\prime)} g^{\mu \alpha} g^{\nu \beta} \right] , \nonumber 
	\\
\end{align}
in which $ t_i $ are the form factors achieved by integrating over the wave functions for specific channels, $ \varepsilon^{\alpha \beta \gamma \delta} $ is the Levi-Civita symbol, $ \epsilon $, $ \epsilon_1 $, and $ \epsilon_2 $ are the polarization tensors or vectors of corresponding states. 
For convenience, we define
\begin{equation}
	\Pi_{\alpha \beta} = -g_{\alpha \beta}+\frac{P_{\alpha} P_{\beta}}{M^{2}},
\end{equation}
where $ P $ and $ M $ are the momentum and mass of the corresponding meson.
Then, the completeness relations of polarization vector and tensor are given by
\begin{equation}
	\begin{aligned}
		\sum_{r} \epsilon^{(r)}_{\mu} \epsilon^{*(r)}_{\mu '} =& \Pi_{\mu \mu '}, \\ 
		\sum_{r} \epsilon^{(r)}_{\mu \nu} \epsilon^{*(r)}_{\mu ' \nu '} =& \frac{1}{2} \left( \Pi_{\mu \mu '} \Pi_{\nu \nu '} + \Pi_{\mu \nu '} \Pi_{\nu \mu '} \right) - \frac{1}{3} \Pi_{\mu \nu} \Pi_{\mu ' \nu '} .
	\end{aligned}
\end{equation}

The two-body decay width can be achieved by  
\begin{equation}
	\begin{split}
		\Gamma=\frac{1}{8 \pi} \frac{\vert \vec{P}_{1} \vert}{M^2} \frac{1}{2J+1}\sum_{r} |\mathcal{M}|^2 ,
	\end{split}
\end{equation}
where $	\vert \vec{P}_{1} \vert = \sqrt{\lambda(M^2,M_1^2,M_2^2)}/2M$ is the momentum of the final charmed meson\footnote{Källén function: $ \lambda (x,y,z) = x^{2}+y^{2}+z^{2}-2xy-2yz-2zx. $}, $ J=2 $ is the spin quantum number of initial $ 2^+ $ state.

\section{Results and Discussions}
\label{sec:results}

In this work, the Cornell potential is taken to solve the BS equations of $ 2^+ $ states numerically.
Within the instantaneous approximation, the Cornell potential in momentum space has the form as follow \cite{Li:2018eqc}
\begin{equation}
	I(\vec{q})=V_{s}(\vec{q})+\gamma_{0} \otimes \gamma^{0} \left[  V_{v}(\vec{q})+V_{0} \right],
\end{equation}
where $ V_0 $ is a free parameter fixed by the physical masses of corresponding mesons; the linear confinement item $ V_s $ and the one-gluon exchange Coulomb-type item $ V_v $ are
\begin{equation}
	\begin{aligned}
		V_s(\vec{q}) &= -\frac{\lambda}{\alpha} (2\pi)^3 \delta^3(\vec{q}) + \frac{8 \pi \lambda}{(\vec{q}^2+\alpha^2)^2} , \\
		V_v(\vec{q}) &= -\frac{16 \pi \alpha_s(\vec{q})}{3(\vec{q}^2+\alpha^2)}.
	\end{aligned}
\end{equation}
The coupling constant $ \alpha_s $ is running
\begin{equation}
	\alpha_s(\vec{q}) = \frac{12\pi}{27}\frac{1}{\ln \left( e + \frac{\vec{q}^2}{\Lambda_{\rm QCD}^2} \right)},
\end{equation}
where $ e = 2.7183 $ .

The parameters adopted in the numerical calculation are listed as follows \cite{Wang:2013lpa}:
\begin{align*}
	&m_u=0.305\GeV,\ m_d=0,311\GeV,\ m_c=1.62\GeV,\\ 
	&m_s=0.500\GeV,\ \alpha=0.060\GeV, \ \lambda=0.210\GeV^2, \\
	& \Lambda_{\rm QCD}=0.270 \GeV, \ f_{\pi}=0.130\GeV, \ f_K=0.156\GeV ,\\
	& f_{\eta_1}=1.07f_{\pi}, \ f_{\eta_8}=1.26f_{\pi},\  M_{\eta_1}=0.923\GeV, \ M_{\eta_8}=0.604\GeV.
\end{align*}
For the doublet of $ 2^- $ charmed states, the masses
$ m_{D_2(2740)}=2.747\GeV$, $ m_{D_2(2780)}=2.780\GeV $ are used. The masses values of other involved mesons are taken from Ref. \cite{ParticleDataGroup:2020ssz}.

Before presenting our results, a remark is in order. When we solve the wave functions, the $ 2^+  $ states actually are the mixture of several partial waves \cite{Wang:2022cxy}. Within this work, to avoid confusion, we will not discuss this mixing in detail and still use pure $ P $ or $ F $ wave to mark each $ 2^+ $ state.

Within our calculation, the wave functions of each state are acquired by fixed on the experimental mass, which also gives the same phase space for every assignment. The total and partial widths of $ D_2^*(3000) $ with possible assignments are presented in Table \ref{tab:width}. 
In the four possible candidates, $ D_2^*(3000) $ as $ 1^3F_2 $ has the largest total width in our prediction, which is about 778.0 MeV and much exceeds the upper limit of present experimental results. The channels of $ D \pi $, $ D^* \pi $, $ D_s K $, $ D_1(2420) /D_1(2430) \pi(\eta) $, $ D_{s1}(2536) /D_{s1}(2460) K $,  and $ D_2(2740)/ D_2(2780) \pi$ contribute much to the total width. In these dominant channels, the branching fraction of most concerned $ D \pi $ mode is about 3\%. And the partial width ratio of $ D^*\pi $ to $ D \pi $ are given by
\begin{equation}
	\begin{aligned}
		\frac{\Gamma [1^3F_2 \to D^*\pi]}{\Gamma [1^3F_2 \to D \pi]} \approx 0.58.
	\end{aligned}
\end{equation}

In the case of $ 2^3P_2 $ state, the total width is estimated to be 285.6 MeV, which is larger than the observational value but still in the error range. The dominant channels include $ D \pi $, $ D^* \pi $, $ D \rho(\omega) $, $ D^* \rho(\omega) $, $ D_s^* K^* $, and $ D_1(2420) /D_1(2430) \pi$. Our predicted branching fractions of $ D\pi $ for $ 2^3P_2 $ is about 4\%. The partial width ratio between $ D^*\pi $ and $ D \pi $ is
\begin{equation}
	\begin{aligned}
		\frac{\Gamma [2^3P_2 \to D^*\pi]}{\Gamma [2^3P_2 \to D \pi]} \approx 0.80.
	\end{aligned}
\end{equation}

The total widths of $ 2^3F_2 $ and $ 3^3P_2 $ are about 61 MeV and 19.1 MeV, respectively. The width of $ 2^3F_2 $ reaches the lower limit while the result of $ 3^3P_2 $ doesn't. The channels of $ D \pi $, $ D^*\pi $, $ D_1(2420)/D_1(2430) \pi$, and $ D_2(2740)/D_2(2780)\pi $ give main contribution to the width of $ 2^3F_2 $, while $ 3^3P_2 $ dominantly decay into $ D\pi $, $ D^*\pi $, $ D^* \rho(\omega) $, and $ D_s^* K^* $. The branching fractions of $ D\pi $ for $ 2^3F_2 $ and $ 3^3P_2 $ are 12\% and 10\%, respectively. The corresponding partial width ratios are
\begin{align}
		\frac{\Gamma [2^3F_2 \to D^*\pi]}{\Gamma [2^3F_2 \to D \pi]} \approx 0.39, \\
		\frac{\Gamma [3^3P_2 \to D^*\pi]}{\Gamma [3^3P_2 \to D \pi]} \approx 1.41.
\end{align}

We notice that $ D\pi $ mode is appreciable for all four candidates, which is consistent with present experimental observations. The widths of channels involving $ D_1(2420) / D_1(2430) $, $ D_{s1}(2536) / D_{s1}(2460) $ and $ D_2(2740) / D_2(2780)$ are also considerable especially for $ 2^3P_2 $, $ 3^3P_2 $ and $ 2^3F_2 $ states. In addition, the partial width ratios of $ D\pi $ to $ D^* \pi $ are different for these candidates, which could be useful for future experimental test.

\begin{table}[htb]
	\footnotesize
	\renewcommand\arraystretch{0.75}
	\caption[results]{The strong decay widths (MeV) of $ D_2^*(3000)^0 $ as different possible assignments. The present experiment mass $ M=3214 \MeV $ is used for each assignment here.}
	\label{tab:width}
	\vspace{0.5em}\centering
	\setlength{\tabcolsep}{0.2em}{
		\begin{tabular}{lcccclcccc}
			\hline
			& $2 ^3P_2$ & $3 ^3P_2$ & $1 ^3F_2$ & $2 ^3F_2$ &                         & $2 ^3P_2$ & $3 ^3P_2$ & $1 ^3F_2$ & $2 ^3F_2$ \\ \hline
			$D^+ \pi^-$                 & 7.19      & 1.28      & 13.9      & 4.85      & $D^{*+} \pi^-$          & 5.78      & 1.81      & 8.02      & 1.91      \\
			$D^0 \pi^0$                 & 3.66      & 0.635     & 6.93      & 2.49      & $D^{*0} \pi^0$          & 2.94      & 0.882     & 4.06      & 0.987     \\
			$D^0 \eta$                  & 0.167     & 0.0258    & 1.81      & 0.520     & $D^{*0} \eta$           & 0.0393    & 0.042     & 0.95      & 0.191     \\
			$D^0 \eta '$                & 0.458     & 0.000679  & 5.01      & 0.590     & $D^{*0} \eta '$         & 0.747     & 0.00699   & 1.46      & 0.139     \\
			$D_0(2550)^+ \pi^-$         & 28.9      & 0.107     & 10.2      & 1.31      & $D_1^{*}(2600)^+ \pi^-$ & 38.3      & 0.0458    & 4.96      & 0.274     \\
			$D_0(2550)^0 \pi^0$         & 14.5      & 0.0565    & 5.17      & 0.681     & $D_1^{*}(2600)^0 \pi^0$ & 19.3      & 0.0245    & 2.54      & 0.144     \\
			$D_0(2550)^0 \eta$          & 0.575     & 0.0418    & 0.262     & 0.0286    & $D_1^{*}(2600)^0 \eta$  & 0.0991    & 0.00298   & 0.0159    & 0.000275  \\
			$D_s^+ K^-$                 & 1.18      & 0.738     & 8.77      & 1.58      & $D_{s}^{*+} K^-$        & 0.762     & 0.841     & 2.97      & 0.509     \\
			&           &           &           &           &                         &           &           &           &           \\
			$D^+ \rho^-$                & 4.76      & 0.879     & 0.753     & 0.458     & $D^{*+} \rho^-$         & 12.5      & 3.40      & 2.61      & 0.690     \\
			$D^0 \rho^0$                & 2.51      & 0.439     & 0.384     & 0.242     & $D^{*0} \rho^0$         & 6.38      & 1.72      & 1.43      & 0.368     \\
			$D^0 \omega^0$              & 2.35      & 0.417     & 0.402     & 0.245     & $D^{*0} \omega^0$       & 6.44      & 1.71      & 1.44      & 0.358     \\
			$D_s^+ K^{*-}$              & 0.381     & 0.315     & 0.306     & 0.0851    & $D_{s}^{*+} K^{*-}$     & 6.88      & 2.62      & 0.118     & 0.0311    \\
			&           &           &           &           &                         &           &           &           &           \\
			$D_0^{*}(2400)^+ \rho^-$    & 5.18      & 0.00184   & 0.251     & 0.00212   & $D_2^{*}(2460)^+ \pi^-$ & 3.29      & 0.0708    & 1.46      & 0.0465    \\
			$D_0^{*}(2400)^0 \rho^0$    & 4.03      & 0.0328    & 0.23      & 0.00503   & $D_2^{*}(2460)^0 \pi^0$ & 1.68      & 0.0331    & 0.743     & 0.0218    \\
			$D_0^{*}(2400)^0 \omega^0$  & 3.87      & 0.0209    & 0.198     & 0.00392   & $D_2^{*}(2460)^0 \eta$  & 0.136     & 0.00494   & 0.102     & 0.00230   \\
			$D_{s0}^{*}(2317)^+ K^{*-}$ & 0.0551    & $ 9.48 \times10^{-5}  $ & 0.00790   & $ 4.29 \times 10^{-5} $  & $D_{s2}(2573)^+ K^-$    & 0.0982    & 0.0165    & 0.140     & 0.0168    \\
			&           &           &           &           &                         &           &           &           &           \\
			$D_1(2420)^+ \pi^-$         & 27.6      & 0.0916    & 75.8      & 8.52      & $D_1(2430)^+ \pi^-$     & 24.7      & 0.340     & 64.9      & 7.40      \\
			$D_1(2420)^0 \pi^0$         & 14.3      & 0.0473    & 38.2      & 4.43      & $D_1(2430)^0 \pi^0$     & 12.8      & 0.192     & 32.8      & 3.90      \\
			$D_1(2420)^0 \eta$          & 0.835     & 0.00784   & 10.1      & 0.695     & $D_1(2430)^0 \eta$      & 0.746     & 0.0174    & 8.70      & 0.595     \\
			$D_{s1}(2536)^+ K^-$        & 1.21      & 0.0105    & 31.8      & 2.27      & $D_{s1}(2460)^+ K^-$    & 3.53      & 0.0646    & 46.0      & 3.93      \\
			&           &           &           &           &                         &           &           &           &           \\
			$D_1(2420)^+ \rho^-$        & 2.34      & 0.0142    & 11.6      & 0.205     & $D_1(2430)^+ \rho^-$    & 0.472     & 0.00360   & 3.07      & 0.0490    \\
			$D_1(2420)^0 \rho^0$        & 1.48      & 0.00887   & 7.32      & 0.128     & $D_1(2430)^0 \rho^0$    & 0.358     & 0.00262   & 2.34      & 0.0358    \\
			$D_1(2420)^0 \omega$        & 0.673     & 0.00572   & 3.41      & 0.0515    & $D_1(2430)^0 \omega$    & 0.0868    & 0.000823  & 0.579     & 0.00784   \\
			&           &           &           &           &                         &           &           &           &           \\
			$D_2(2740)^+ \pi^-$         & 3.78      & 0.0223    & 123       & 3.60      & $D_2(2780)^+ \pi^-$     & 2.42      & 0.0461    & 113       & 2.70      \\
			$D_2(2740)^0 \pi^0$         & 1.91      & 0.00914   & 61.2      & 1.84      & $D_2(2780)^0 \pi^0$     & 1.22      & 0.0229    & 56.6      & 1.39      \\ \hline
			Total                       & 285.6     & 19.1      & 778.0     & 60.5      &                         &           &           &           &           \\ \hline
	\end{tabular}}
\end{table}

\begin{figure}[htb!]
	\centering
	\subfigure[$ D_2^*(3000)^0 $ as $ 2^3P_2 $ state]{
		\label{angle_1+_2P} 
		\includegraphics[width=0.48\textwidth]{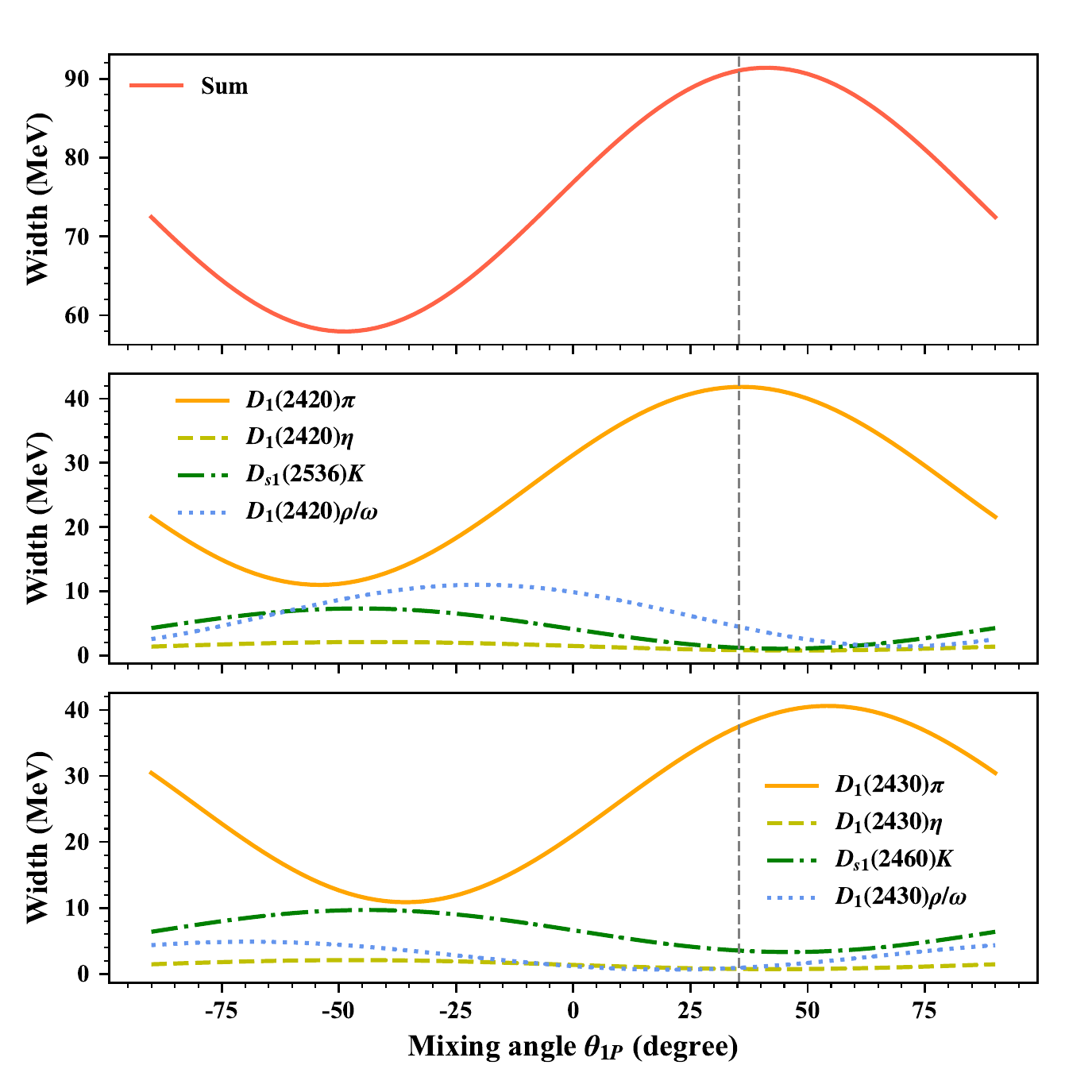}}
	\subfigure[$ D_2^*(3000)^0 $ as $ 3^3P_2 $ state]{
		\label{angle_1+_3P} 
		\includegraphics[width=0.48\textwidth]{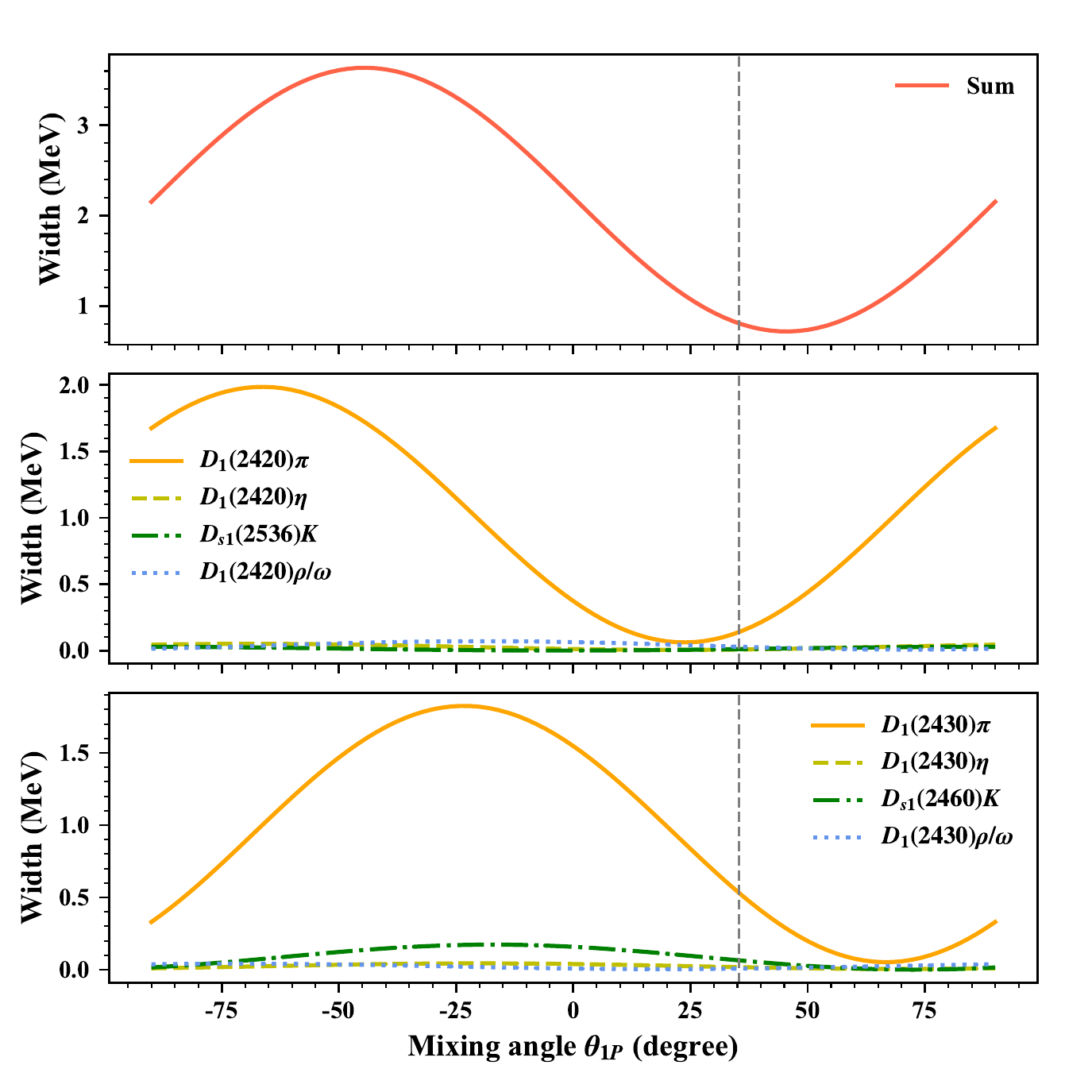}}
	
	\subfigure[$ D_2^*(3000)^0 $ as $ 1^3F_2 $ state]{
		\label{angle_1+_1F} 
		\includegraphics[width=0.48\textwidth]{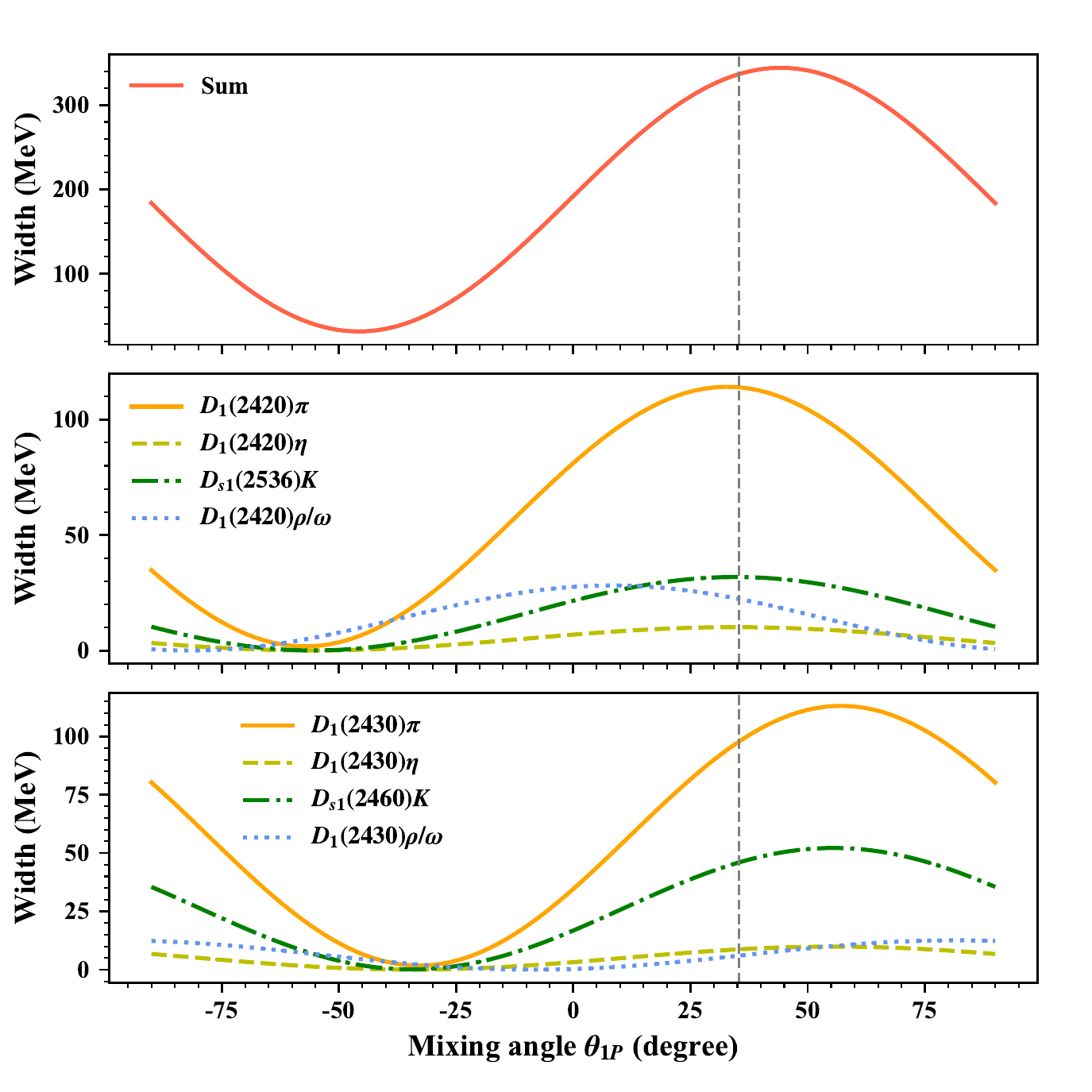}}
	\subfigure[$ D_2^*(3000)^0 $ as $ 2^3F_2 $ state]{
		\label{angle_1+_2F} 
		\includegraphics[width=0.48\textwidth]{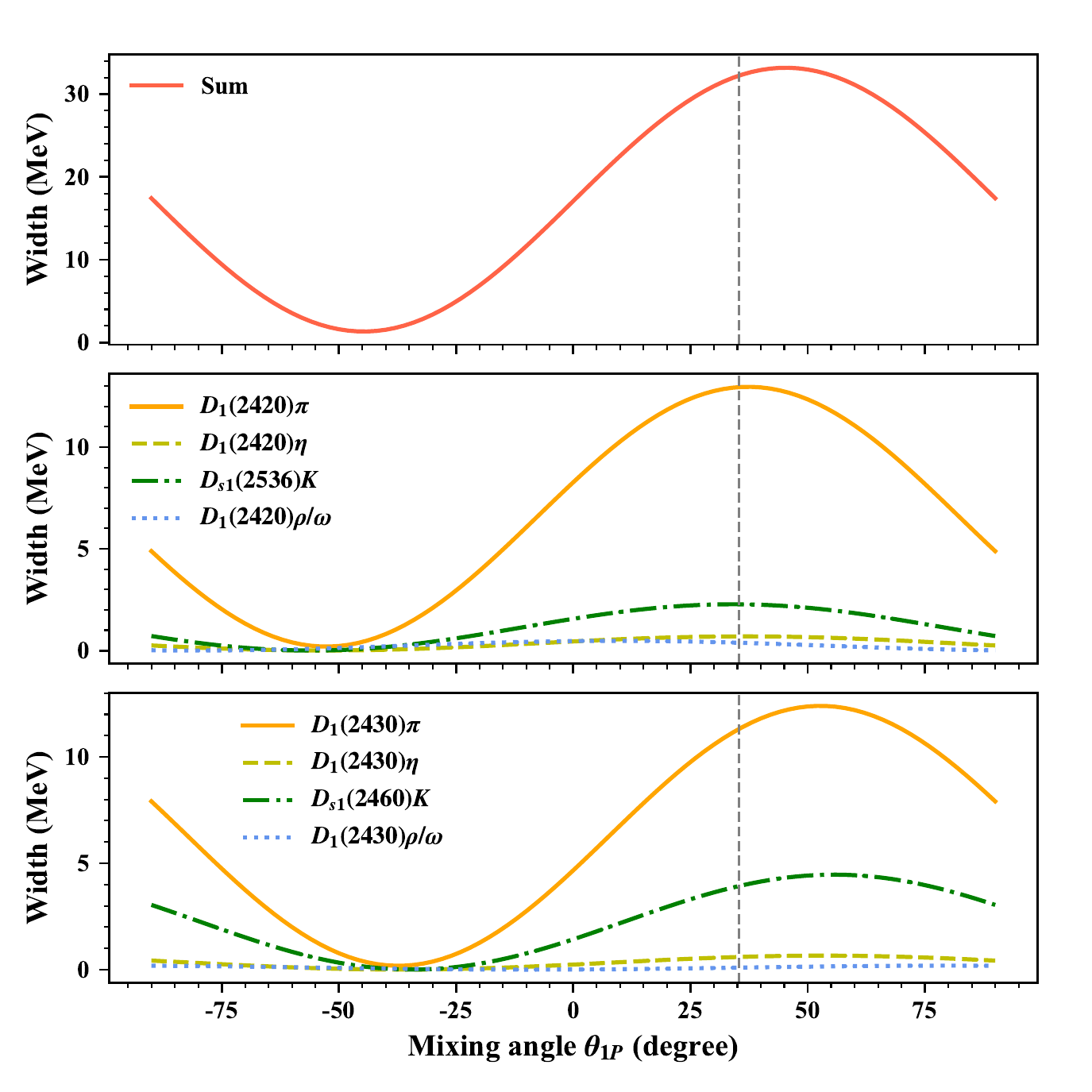}}
	\caption{The dependence of partial widths of channels involving final $ D_1(2420)/D_{s1}(2536) $ or $ D_1(2430)/D_{s1}(2460) $ mesons on mixing angle $\theta_{1P}$. The vertical dashed line indicates the ideal mixing angle $ \theta_{1P}=\arctan(\sqrt{1/2}) \approx 35.3 \degree$.}
	\label{fig:angle-runing_1+}
\end{figure}

\begin{figure}[htb!]
	\centering
	\subfigure[$ D_2^*(3000)^0 $ as $ 2^3P_2 $ state]{
		\label{angle_2-_2P} 
		\includegraphics[width=0.48\textwidth]{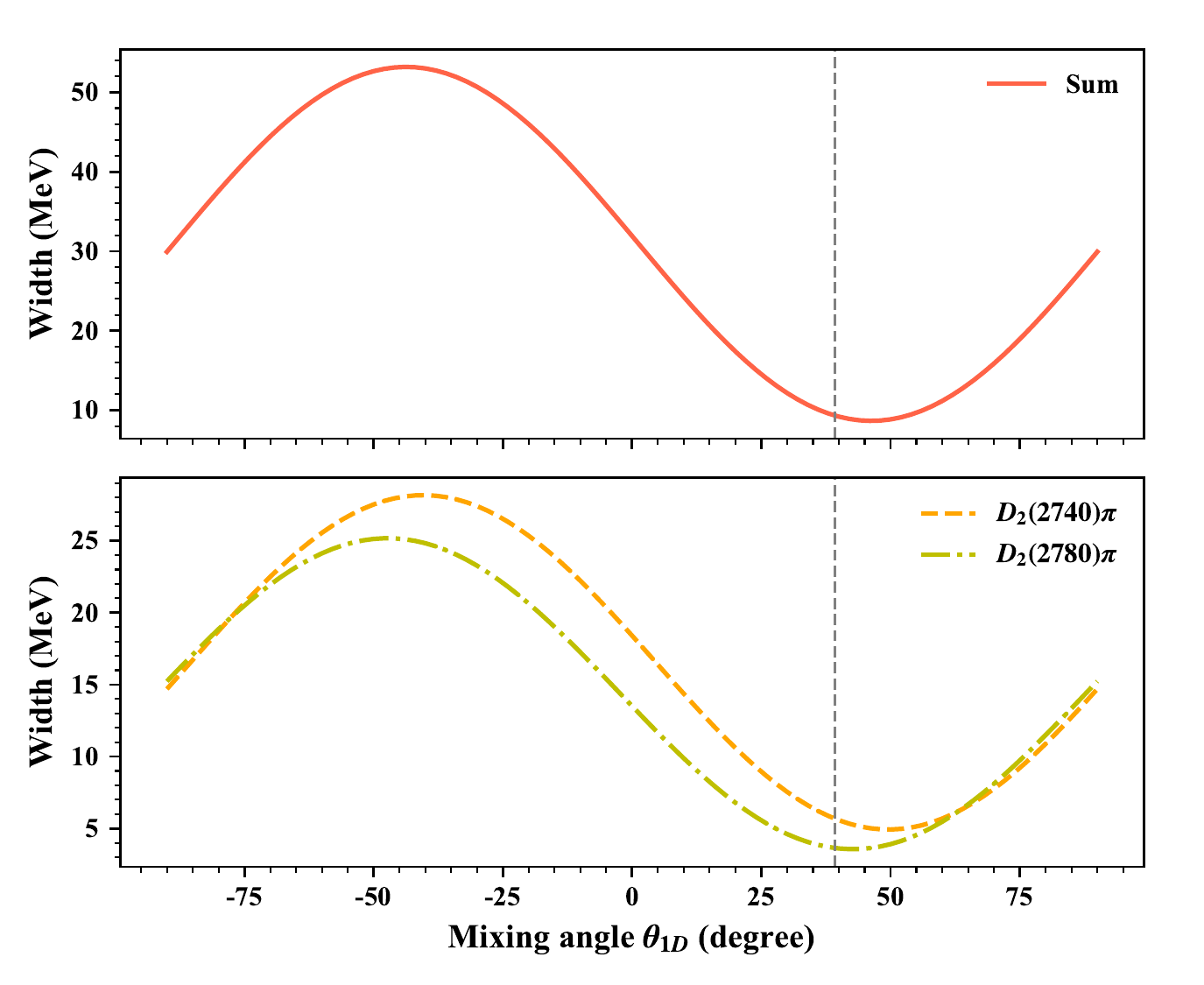}}
	\subfigure[$ D_2^*(3000)^0 $ as $ 3^3P_2 $ state]{
		\label{angle_2-_3P} 
		\includegraphics[width=0.48\textwidth]{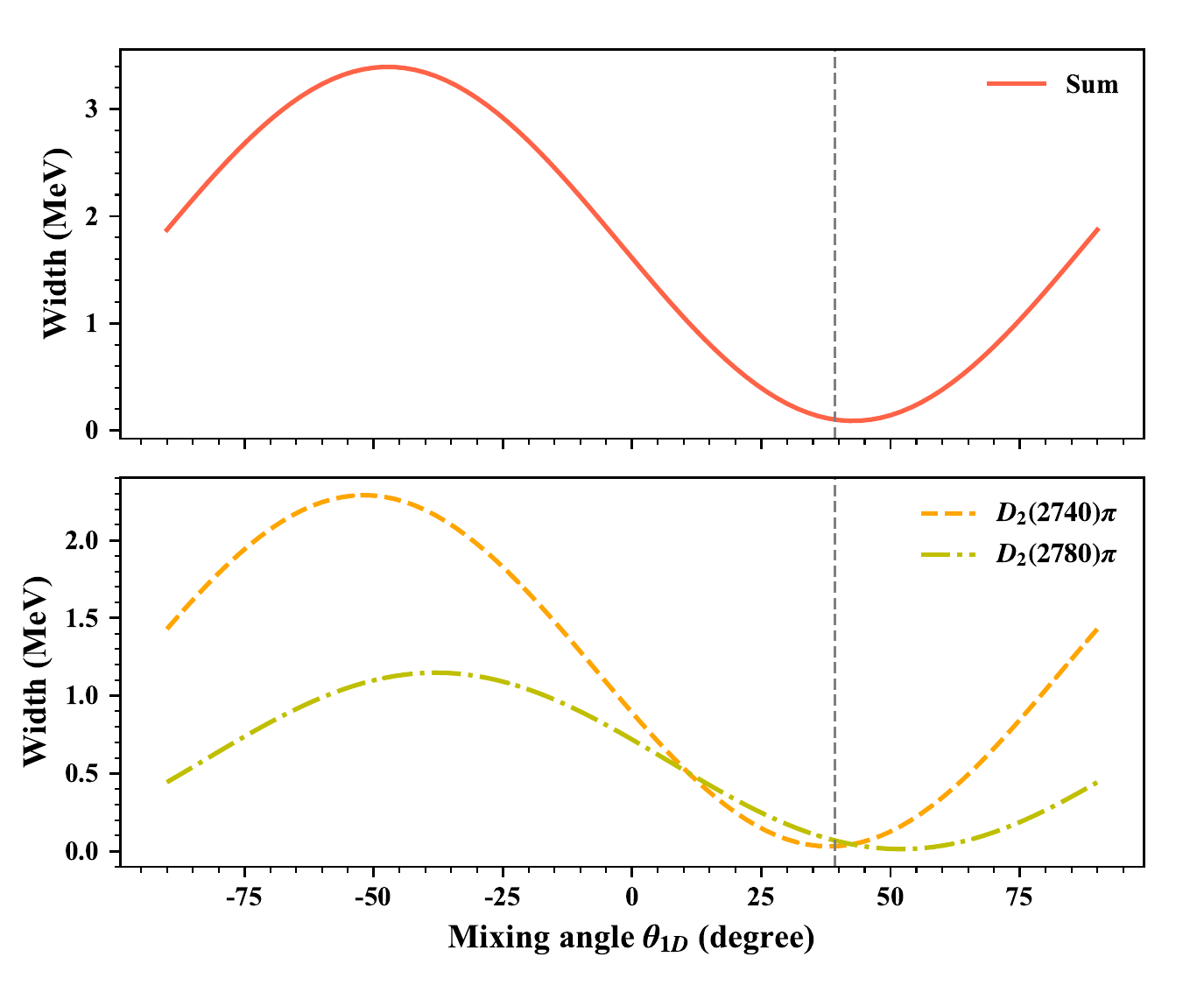}}
	
	\subfigure[$ D_2^*(3000)^0 $ as $ 1^3F_2 $ state]{
		\label{angle_2-_1F} 
		\includegraphics[width=0.48\textwidth]{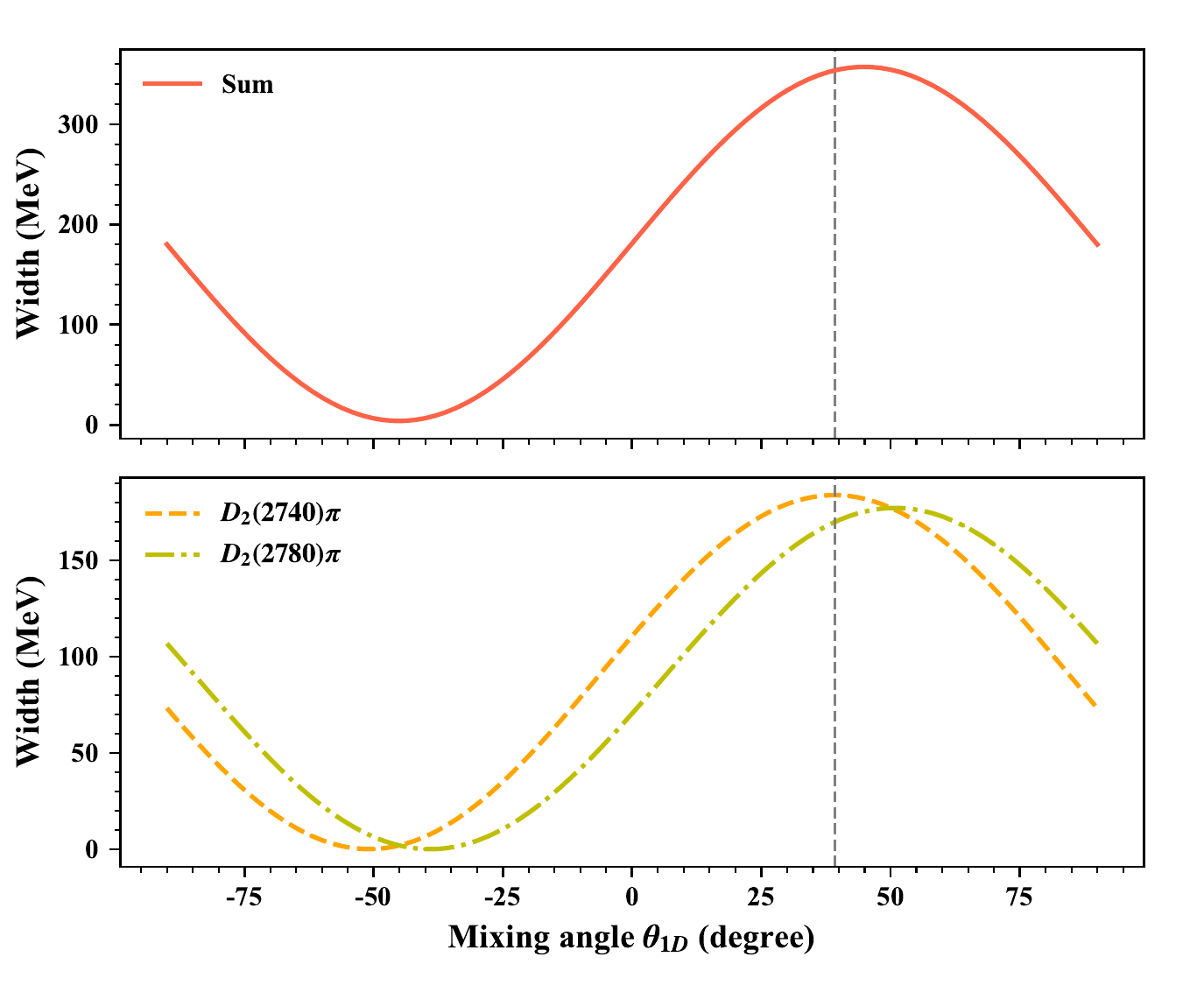}}
	\subfigure[$ D_2^*(3000)^0 $ as $ 2^3F_2 $ state]{
		\label{angle_2-_2F} 
		\includegraphics[width=0.48\textwidth]{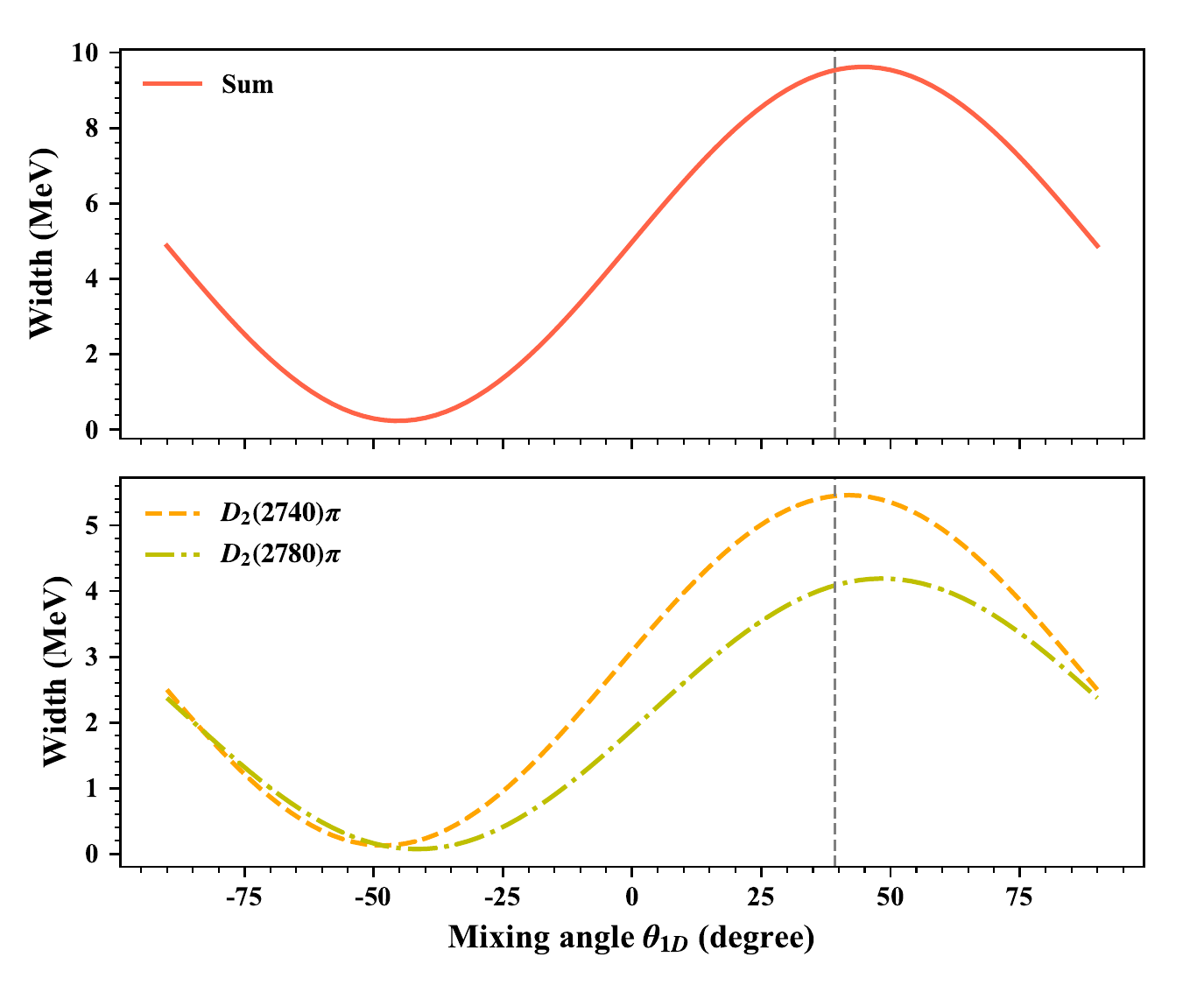}}
	\caption{ The dependence of partial widths of channels involving final $ D_2(2740) $ or $ D_2(2780)$ mesons on mixing angle $\theta_{1D}$. The vertical dashed line indicates the ideal mixing angle $ \theta_{1D}=\arctan(\sqrt{2/3}) \approx 39.2\degree $.}
	\label{fig:angle-runing_2-}
\end{figure}

The results in Table \ref{tab:width} are calculated by using ideal mixing and the divergence of mixing angle could give large corrections. Thus we show the dependence of partial widths on mixing angle $ \theta_{1P}$ and $\theta_{1D} $ in Fig. \ref{fig:angle-runing_1+} and \ref{fig:angle-runing_2-}. As $ 1^3F_2 $ and $ 2^3F_2 $ states, the widths could reduce by up to about 640 MeV and 40 MeV in total, respectively if choosing the negative angles for both $ \theta_{1P}$ and $\theta_{1D} $. In the instance of $ 2^3P_2 $ and $ 3^3P_2 $ states, the total widths could approximately increase by 10 MeV and 6 MeV, respectively when the negative mixing angles are appointed. 
The mixing has been discussed by our previous works \cite{Li:2018eqc,Wang:2018psi} and ideal mixing is valid for $D_1(2420)$ and $D_1(2430)$. Here we only show the sensitive dependence between partial widths and mixing angle. In the following discussion, we keep using the results of ideal mixing for consistency.

\begin{figure}[htb!]
	\centering
	\subfigure[$ D_2^*(3000)^0 $ as $ 2^3P_2 $]{
		\label{BSE_2P} 
		\includegraphics[width=0.48\textwidth]{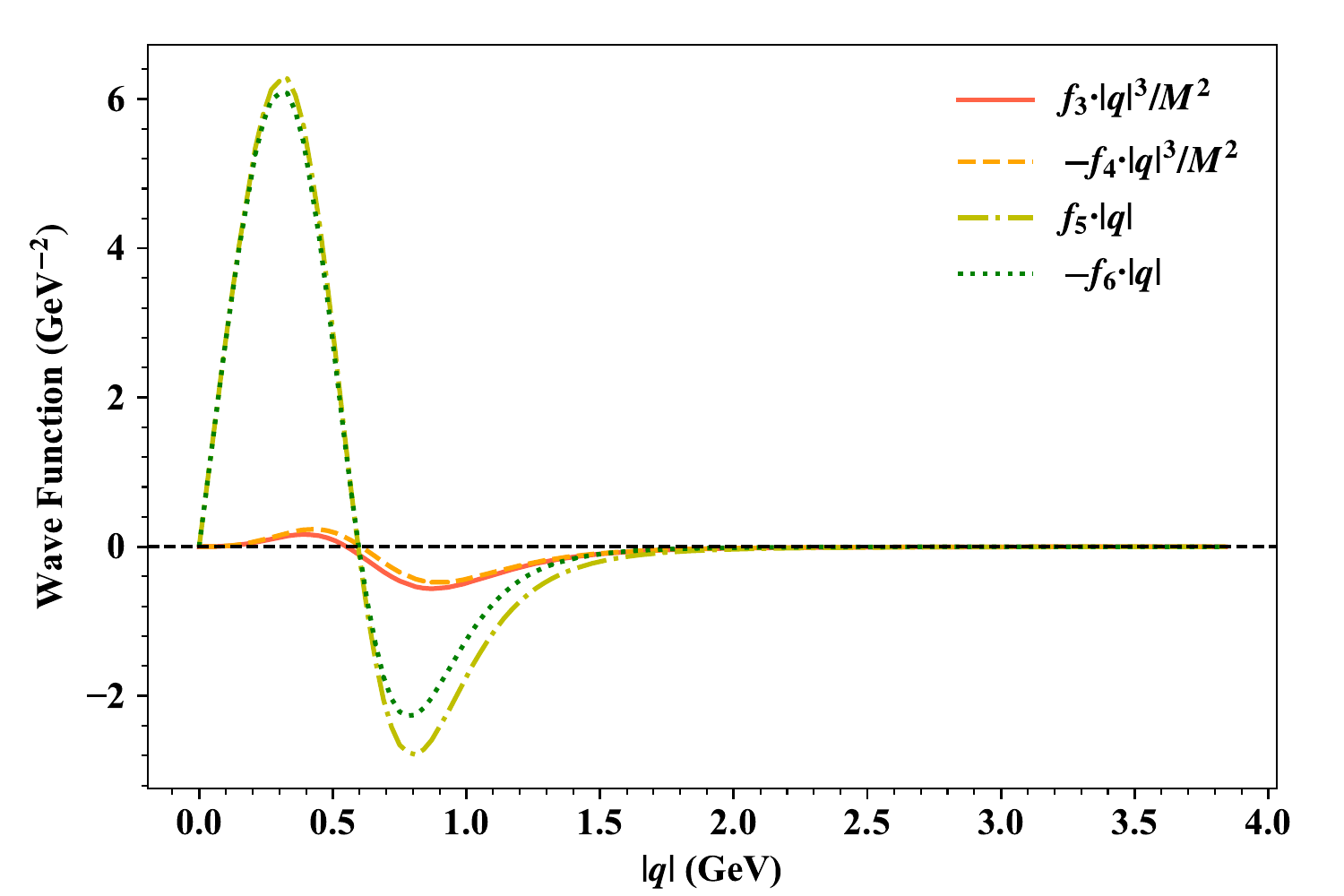}}
	\subfigure[$ D_2^*(3000)^0 $ as $ 3^3P_2 $]{
		\label{BSE_3P} 
		\includegraphics[width=0.48\textwidth]{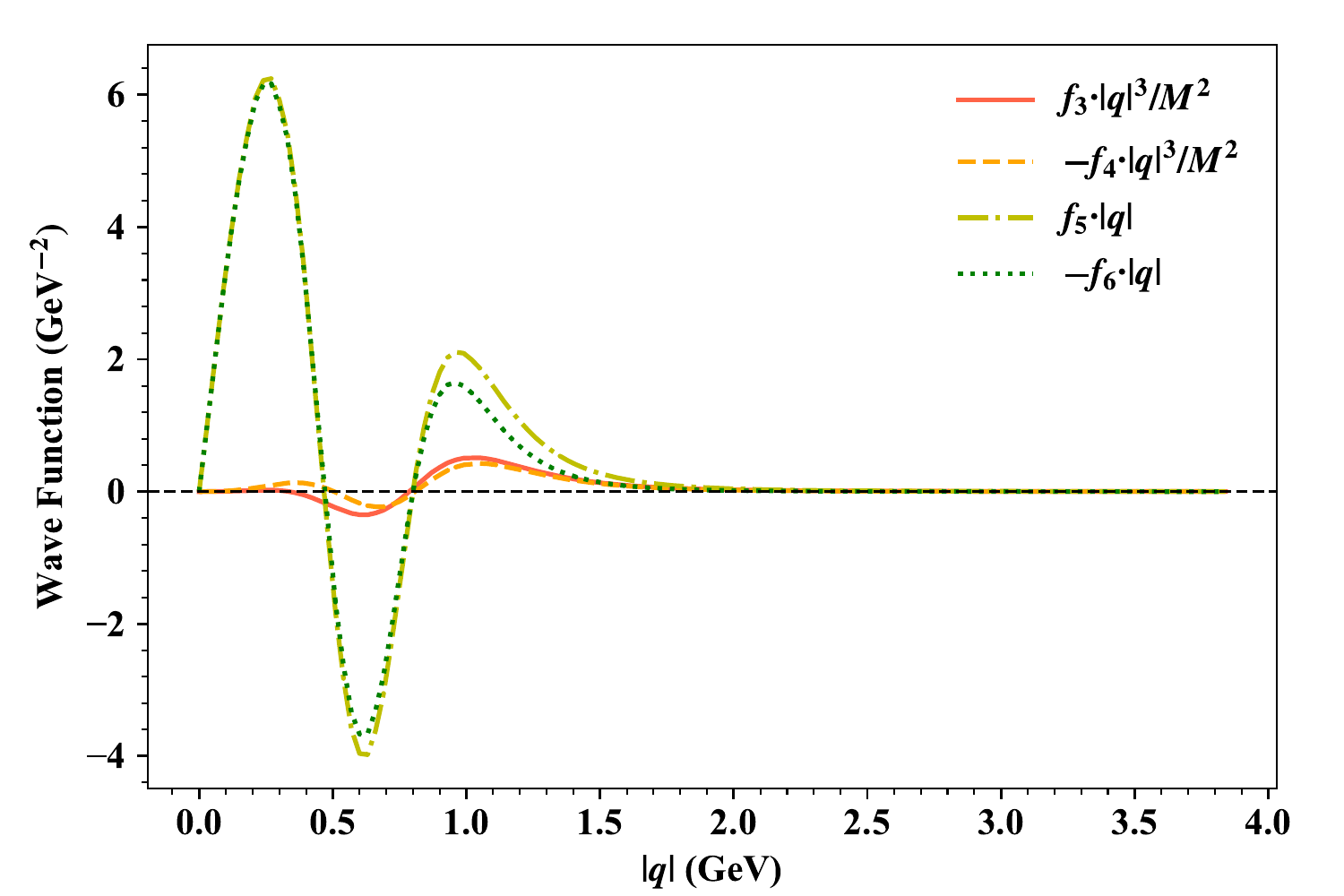}}
	
	\subfigure[$ D_2^*(3000)^0 $ as $ 1^3F_2 $]{
		\label{BSE_1F} 
		\includegraphics[width=0.48\textwidth]{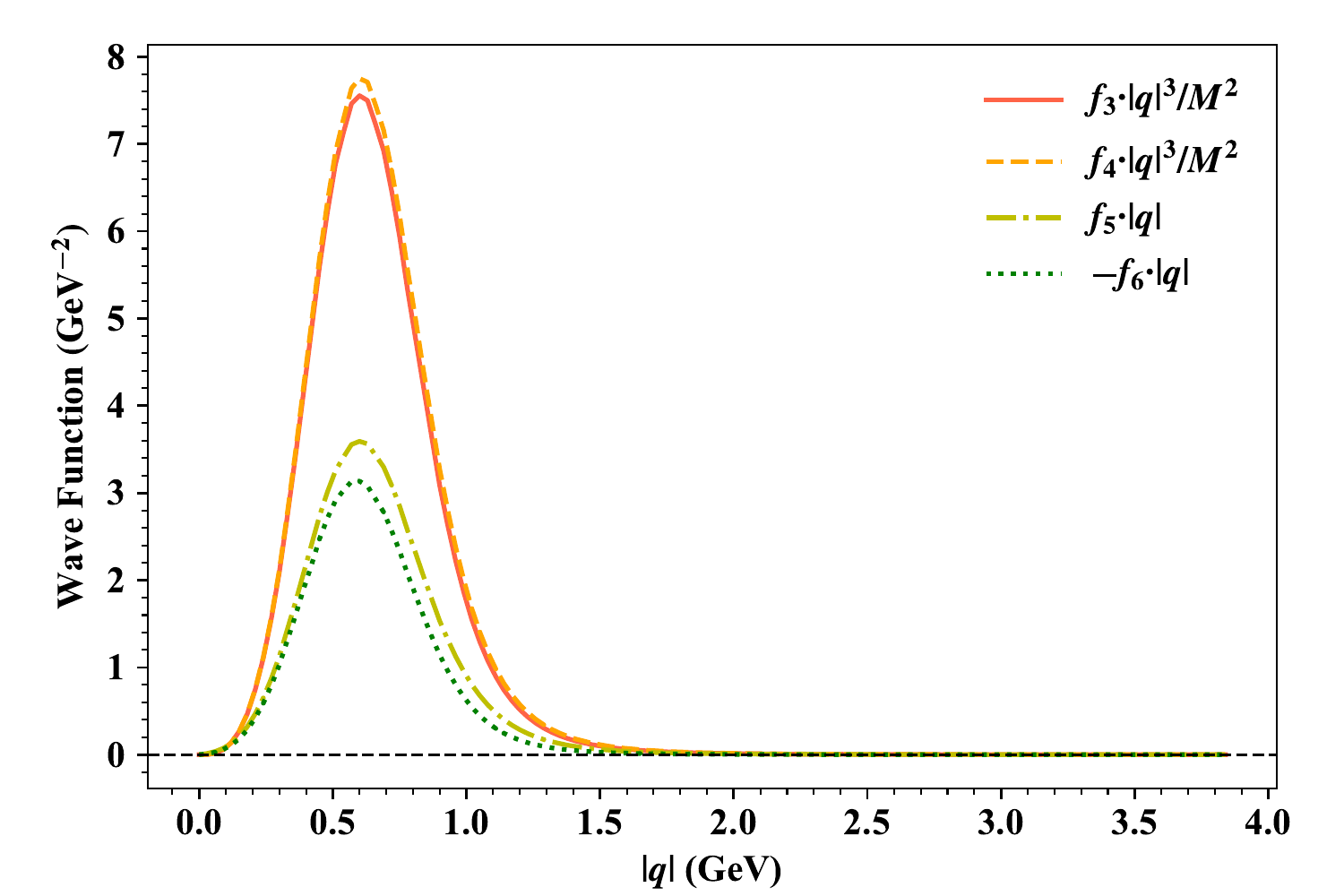}}
	\subfigure[$ D_2^*(3000)^0 $ as $ 2^3F_2 $]{
		\label{BSE_2F} 
		\includegraphics[width=0.48\textwidth]{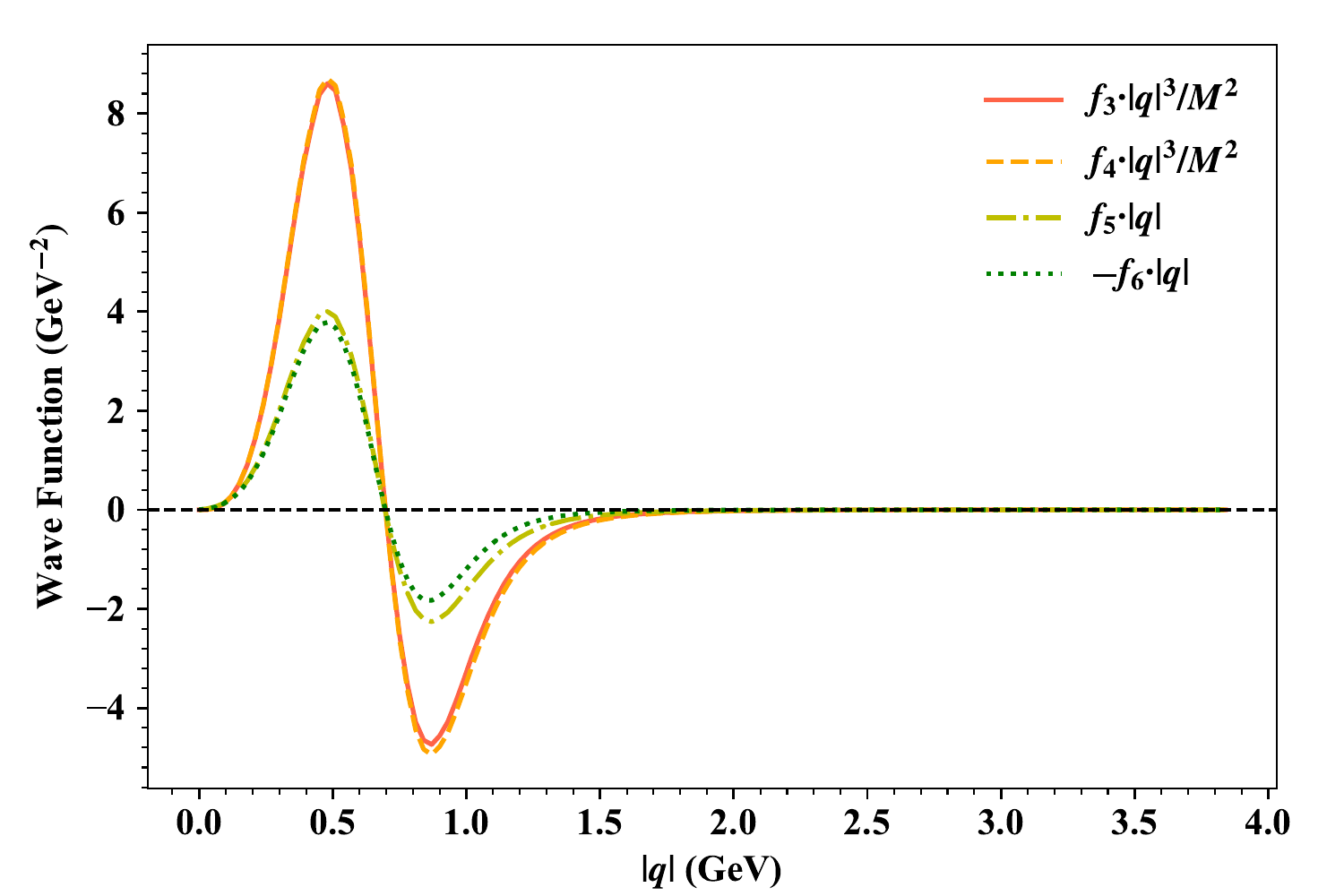}}
	
	\subfigure[$ D^0 $ as $ 1^1S_0 $]{
		\label{BSE_1S} 
		\includegraphics[width=0.48\textwidth]{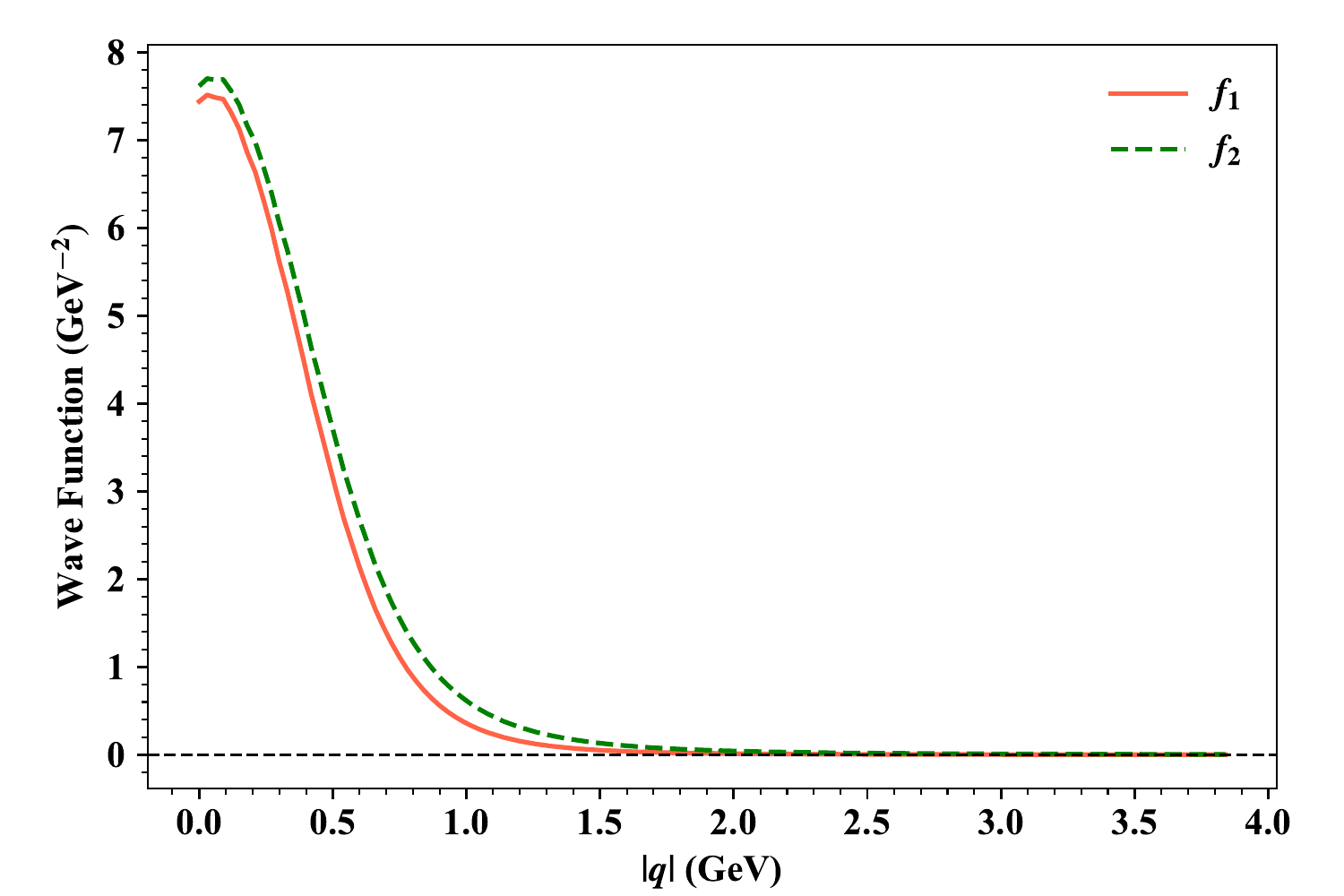}}
	\subfigure[$ D_0(2550)^0 $ as $ 2^1S_0 $]{
		\label{BSE_2S} 
		\includegraphics[width=0.48\textwidth]{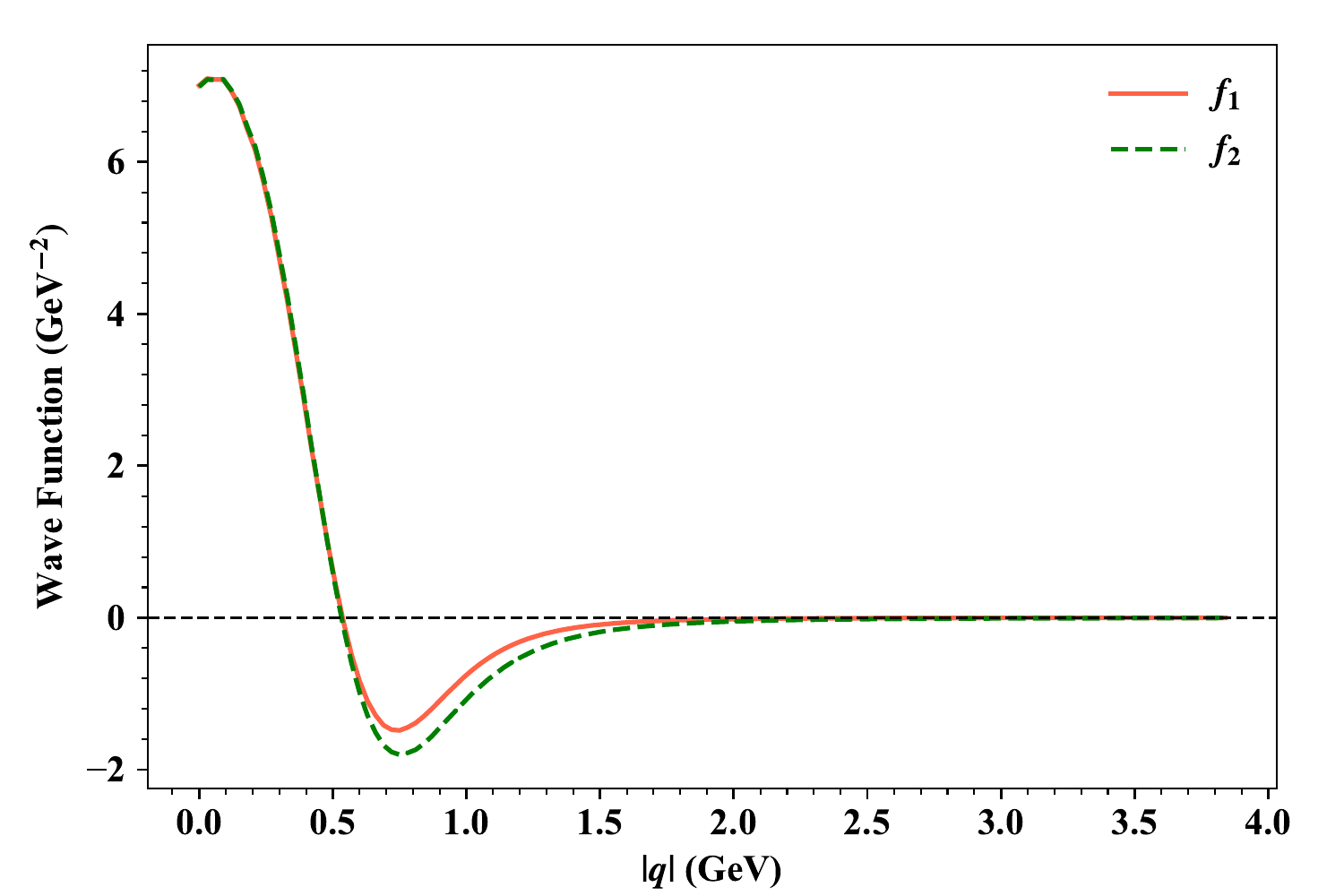}}
	\caption{Some wave functions of $ D_2^*(3000) $ as different assignments, $ D^0 $ as $ 1 ^1S_0 $, and $ D_0(2550)^0 $ as $ 2 ^1S_0 $.}
	\label{fig:wavefunction}
\end{figure}

Although all four assignments are calculated by fitting the same mass value $ M = 3214 \MeV $, which gives equivalent phase space for the same channels, the decay behaviors of these states are quite different.
In our work, the discrepancy of total widths for different assignments can be explained by the special structure of wave functions.
The numerical wave functions of $ D_2^*(3000) $ with different states and $ D(1^1S_0)/D(2^1S_0) $ as an example in final states are shown as Fig. \ref{fig:wavefunction}. $ 2^3P_2 $, $ 3^3P_2 $, $ 2^3F_2 $ and $ 2^1S_0 $ states have nodes and the wave function values alter the sign after nodes, which could lead to cancellation (or enhancement) in the overlapping integration. Such as, the width of $ D(2^3P_2) \to D(2^1S_0)\pi $ is larger than $ D(2^3P_2) \to D(1^1S_0)\pi $. Both $ 2^3P_2 $ and $ 2^1S_0 $ have one node, which finally enhance the integral. However, in the case of $ 3^3P_2 $, the structure of two nodes eventually gives the contrary behavior with same two channels. 
As for $ 1^3F_2 $ and $ 2^3F_2 $, since the shape of wave functions are different ($ f_3 $, $ f_4 $, and $ f_5 $ always have the same sign for $ n^3F_2 $, while $ f_3 $ and $ f_4$ have the same sign for $ n^3P_2 $), the widths of $ D(2^1S_0)\pi $ are smaller than $ D(1^1S_0)\pi $. In general, $ 1^3F_2 $ and $ 2^3P_2 $ states reach larger widths with less cancellation, while $ 2^3F_2 $ and $ 3^3P_2 $ behave oppositely. It also illustrate that the correction from higher internal momentum $ q $ of mesons is non-negligible.

\begin{figure}[htb!]
	\centering
	\includegraphics[width=0.7\textwidth]{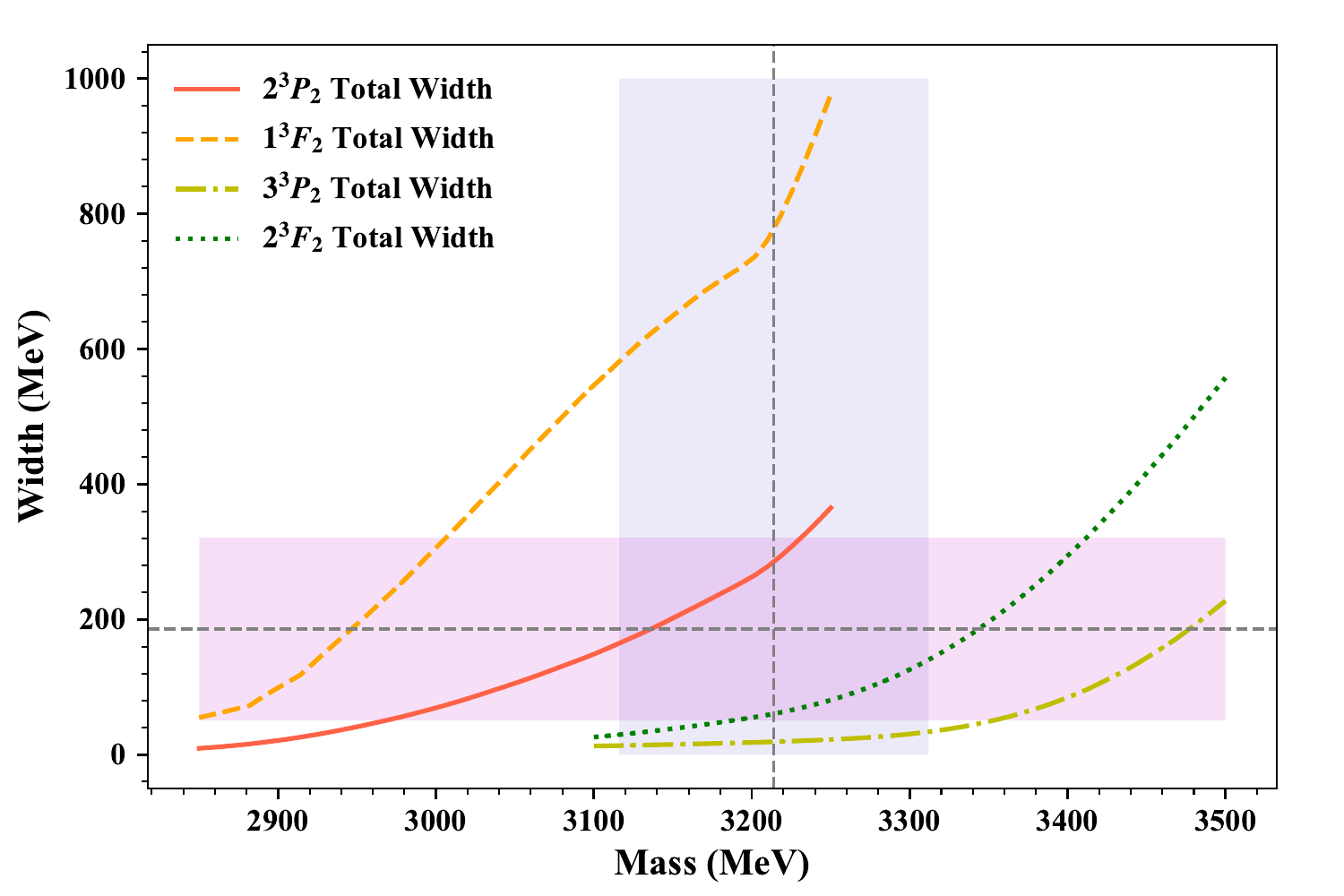}
	\caption{Total widths change with the masses of different states for $ D_2^* $. The vertical \& horizon dashed lines with shades are present experimental results with errors \cite{LHCb:2016lxy}.}
	\label{fig:mass-running}
\end{figure}

At last, we list the full widths of excited $ 2^+ $ states from some other models in Table \ref{tab:refwidth} for comparison. Different methods give inconsistent results at present because the mass values used in these references are various. Considering the predicted mass spectrum of $ 2^+ $ charmed states, we show our total widths of $ 2^3P_2$, $ 1^3F_2 $, $ 3^3P_2$, and $ 2^3F_2 $ change with masses in the range of $ 2850 \sim 3250 \MeV $ and $ 3100 \sim 3500 \MeV $, respectively in Fig. \ref{fig:mass-running}. We can find that the widths vary widely along with the masses. 
At $ M = 2900 \sim 3000 \MeV $ and $ M = 3050 \sim 3150 \MeV $,  $ 2^3P_2 $ and $ 1^3F_2 $ states reach the total width of $ 21 \sim 69 \MeV  $ and $ 427 \sim 649 \MeV $ subsequently, which shows an similar trend with the results of Ref. \cite{Godfrey:2015dva} and \cite{Song:2015fha}. In like manner, the widths of $ 3^3P_2 $ and $ 2^3F_2 $ states are in the range of $ 22 \sim 49 \MeV $ and $ 196 \sim 416 \MeV $ at the mass of $ M = 3250 \sim 3350 \MeV $ and $ M = 3350 \sim 3450 \MeV $, which is very roughly close to the value of Ref.\cite{Godfrey:2015dva} and \cite{Wang:2016krl}, respectively. 
At last, when comparing our results with current experimental observation\cite{LHCb:2016lxy}, $ 2^3P_2 $ and $ 2^3F_2 $ states enter the error range which is denoted as the shade area.

\begin{table}[htb]
	\renewcommand\arraystretch{1}
	\caption[results]{Different theoretical predicted widths (MeV) for charmed $2^+$ states. The values in parentheses are the masses used in the corresponding references. Others without parentheses use $ M = 3214 \MeV $. }
	\label{tab:refwidth}
	\setlength{\tabcolsep}{3pt}
	\vspace{0.5em}\centering
	\begin{tabular}{lcccccccc}
		\hline
		& Sun\cite{Sun:2013qca} & Xiao\cite{Xiao:2014ura} & Godfrey\cite{Godfrey:2015dva} & Song\cite{Song:2015fha} & Yu\cite{Yu:2016mez} & Wang\cite{Wang:2016krl} & Ni\cite{Ni:2021pce} & Ours   \\ \hline
		$2 ^3P_2$ & 47(3008)              & 150(3020)               & 114(2957)                     & 68.89(2884)             & 442.36              & -                       & 193.4(2955)         & 285.57 \\
		$1 ^3F_2$ & 136(3008)             & 900(3100)               & 243(3132)                     & 222.02(3053)            & 220.05              & -                       & 722(3096)           & 778.10 \\
		$3 ^3P_2$ & -                     & -                       & 116(3353)                     & -                       & 62.57               & 102.4(3234)             & -                   & 19.13  \\
		$2 ^3F_2$ & -                     & -                       & 223(3490)                     & -                       & 32.09               & 302.2(3364)             & -                   & 60.53  \\ \hline
	\end{tabular}
\end{table} 

\section{Summary}
\label{sec:summary}

We analyzed the OZI-allowed two-body strong decay behaviors of four possible assignments of $ 2^+ $ family for the newly observed $ D_2^*(3000) $. The wave functions of related charmed mesons was obtained by using the instantaneous Bethe-Salpeter method. In our calculation, $ D_2^*(3000) $ as $ 1^3F_2 $ state has an exceeded width of 778 MeV, while the widths of 19.1 MeV for $ 3^3P_2 $ doesn't reach the lower limit. As for $ 2^3P_2 $ and $ 2^3F_2 $, the total widths are respectively 285.6 MeV and 60.5 MeV, which enter the error range of present experimental results. $ D\pi $ channels are important for all four candidates, which shows consistency with the current experiment. Additionally, the channels involving $ D_1(2420)/D_1(2430) $, $ D_{s1}(2536)/D_{s1}(2460) $, and $ D_2(2740) / D_2(2780)$ also give significant contribution and the influence of mixing angle was discussed.
Our study also indicates a strong dependence between total widths and states masses. Thus different models with varying mass input don't reach a consensus. Considering the large uncertainties of preliminary observation, besides the possible individual states, mixing of several states is also a potential option for the current $ D_2^*(3000) $. More accurate measurements and theoretical efforts are expected in the future.

\section*{Acknowledgments }

This work was supported in part by the National Natural Science Foundation of China (NSFC) under Grant No. 12075073 and No. 12005169, the  Natural Science Foundation of Hebei province under the Grant No.  A2021201009, and the Natural Science Basic Research Program of Shaanxi under the Grant No. 2021JQ-074. We also thank the HEPC Studio at Physics
School of Harbin Institute of Technology for access to computing
resources through INSPUR-HPC@hepc.hit.edu.cn.

\bibliography{paper.bib}


\end{document}